\documentstyle[11pt,epsfig]{article}
\title{\bf A cosmological viewpoint on the correspondence between deformed phase-space and canonical quantization}
\author{N. Khosravi$^{1,2,}$\thanks{email: nima@ipm.ir, n-khosravi@sbu.ac.ir}, H. R. Sepangi$^{2,}$\thanks{email: hr-sepangi@sbu.ac.ir} and
 B. Vakili$^{3,}$\thanks{email: b-vakili@cc.sbu.ac.ir, bvakili45@gmail.com}\\\\
$^1${{\small \it{ School of Physics, Institute for Research in
Fundamental Sciences (IPM), P.O.Box 19395-5531, Tehran, Iran}}}\\
$^2${\small {\it Department of Physics, Shahid Beheshti University,
G. C., Evin,
Tehran 19839, Iran }} \\
$^3${\small {\it Department of Physics, Azad University of Chalous,
P. O. Box 46615-397, Chalous, Iran}}}

\begin{document}
\maketitle
\begin{abstract}
We employ the familiar canonical quantization procedure in a given
cosmological setting to argue that it is equivalent to and results
in the same physical picture if one considers the deformation of the
phase-space instead. To show this we use a Probabilistic
Evolutionary Process (PEP) to make the solutions of these different
approaches comparable. Specific model theories are used to show that
the independent solutions of the resulting Wheeler-DeWitt  equation
are equivalent to solutions of the deformation method with different
signs for the deformation parameter. We also argued that since the
Wheeler-DeWitt equation is a direct consequence of diffeomorphism
invariance, this equivalence is only true provided that the
deformation of phase-space does not break such an
invariance.\vspace{5mm}\newline PACS: 98.80.Qc
\end{abstract}
\maketitle
\section{Introduction}
Standard cosmological models based on classical general relativity
have no convincing and precise answer for the presence of the
so-called ``Big-Bang'' singularity. Any hope of dealing with such
singularities would be in vein unless a reliable quantum theory of
gravity can be constructed. In the absence of a full theory of
quantum gravity, it would be useful to describe the quantum states
of the universe within the context of quantum cosmology, introduced
in the works of DeWitt \cite{1} and later Misner \cite{2}. In this
formalism which is based on the canonical quantization procedure,
one first freezes a large number of degrees of freedom and then
quantizes the remaining ones. The quantum state of the universe so
obtained is then described by a wave function in the
mini-superspace, a function of the 3-geometry of the model and
matter fields presented in the theory, satisfying the Wheeler-DeWitt
(WD) equation. In more recent times, the research in this area has
been quite active with different approaches \cite{3}, see \cite{4}
for a review. Interesting applications can be found in \cite{5}
where canonical quantization is applied to many models with
different matter fields as the source of gravity.

An important ingredient in any model theory related to the
quantization of a cosmological model is the choice of the
quantization procedure used to quantized the system. As mentioned
above, the most widely used method has traditionally been the
canonical quantization method based on the WD equation which is
nothing but the application of the Hamiltonian constraint to the
wave function of the universe. However, one may solve the constraint
before using it in the theory and in particular, before quantizing
the system. If we do so, we are led to a Schr\"{o}dinger type
equation where a time reparameterization in terms of various
dynamical variables can be done before quantization \cite{6,7,8}. A
particularly interesting but rarely used approach to study quantum
effects is to introduce a deformation in the phase-space of the
system. It is believed that such a deformation of phase-space is an
equivalent path to quantization, in particular to canonical and path
integral quantizations \cite{quantization}.

An important question then naturally arises in applying the various
quantization methods to cosmological models, that is, if they are
equivalent. To at least partially answer the above question, we
propose to quantize some simple cosmological models, namely the de
Sitter, dusty FRW, FRW with radiative matter and Bianchi type I,
using the methods described above. First we introduce a deformation
to the Poisson algebra of the corresponding phase-space of the
models. This will lead us to a deformed Hamiltonian from which the
equations of motion can be constructed. We will show that the
presence of the deformation parameter in the solutions can be
interpreted as a quantum effect. This is done by comparing the
resulting solutions with that of the other quantization method which
is nothing but the usual canonical WD approach. Indeed, we will show
that both of these quantization methods have the same physical
interpretation for our chosen models.

A remark on the WD approach is that, as is well known, in canonical
quantization the evolution of states disappear since in all
diffeomorphism invariant theories the Hamiltonian becomes a
constraint. Indeed, the wave function in the WD equation is
independent of time, {\it i.e.} the universe has a static picture in
this scenario. On the other hand in the deformed phase-space method,
time does appear and so the evolution becomes meaningful. Therefore,
the problem of the evolution of states which is a major problem in
quantum cosmology may also be addressed in this approach. To propose
a possible solution to this problem we will use what we call the
``Probabilistic Evolutionary Process'' which was introduced in
\cite{nimagrg,grgg}. This mechanism is briefly studied in the next
section because of its crucial role in this work. In section \ref{3}
we review the quantization of the de Sitter and dusty FRW
cosmological models inspired by the Deformed Special Relativity
(DSR), discussed in \cite{jcap,pla}. The canonical quantization for
both models will be studied in section \ref{ca}. We will close the
paper with a discussion and comparison of the results.
\section{Probabilistic Evolutionary Process (PEP)}
To quantize a classical model the following procedure is usually
followed. The classical Hamiltonian is written in its corresponding
operator form and the resulting Schr\"{o}dinger equation obtained
after quantization, {\it i.e.} $i\hbar \frac{\partial}{\partial
t}\Psi={\cal{H}}\Psi$, becomes the relevant equation to describe
time evolution of the quantum states. However, in diffeomorphism
invariant models the Hamiltonian becomes a constraint,
${\cal{H}}=0$, and therefore does not provide for the evolution of
the correspond states. This means that in such models all the states
are stationary. One of the common examples of this situation is
general relativity which is employed to investigate the evolution of
the cosmos. In quantum cosmology, the Schr\"{o}dinger equation
becomes the Wheeler-DeWitt equation, ${\cal{H}}\Psi=0$. As is well
known, quantum cosmology suffers from a number of problems, namely
the construction of the Hilbert space for defining a positive
definite inner product of the solutions of the WD equation, the
operator ordering problem and most importantly, the problem of time;
the wave function in the WD equation is independent of time, {\it
i.e.} the universe has a static picture in this scenario. This
problem was first addressed in \cite{1} by DeWitt himself. However,
he argued that the problem of time should not be considered as a
hinderance in the sense that the theory itself must include a
suitable well-defined time in terms of its geometry or matter
fields. In this scheme time is identified with one of the characters
of the geometry, usually the scale factor and is referred to as the
intrinsic time, or with the momentum conjugate to the scale factor
or even with a scalar character of the matter fields coupled to
gravity in any specific model, known as the extrinsic time.

In general, the crucial problem in canonical quantum gravity is the
presence of constraints in the gravitational field equations.
Identification of time with one of the dynamical variables depends
on the method we use to deal with theses constraints. Different
approaches arising from these methods have been investigated in
details in \cite{6}. As discussed in \cite{6}, time may be
identified before or after quantization has been done. There are
approaches, on the other hand, in which time has no fundamental
role. The problem is how one can describe the evolution of the
universe since observations show that the universe is not presently
in a stationary state. In a previous work \cite{nimagrg,grgg}, we
introduced a mechanism which we have called the Probabilistic
Evolutionary Process (PEP), based on the probabilistic structure of
quantum systems, to provide a sense of the evolution embedded in the
wave function of the universe. This is based on the fact that in
quantum systems the square of a state defines the probability,
${\cal{P}}_a=|\Psi(a)|^2$. The mechanism introduced as PEP says that
the state $\Psi_a$ makes a transition to the state $\Psi_{a+da}$ if
their distance, $da$, is infinitesimal and continuous, and that the
higher the value of the transition probability\footnote{This
transition probability can play the role of the speed of transition
{\it i.e.} the higher the probability of transition the larger the
speed of transition.} the larger the value of
${\cal{P}}_{a+da}-{\cal{P}}_{a}$\footnote{Since there is no
constraint on the positivity of ${\cal{P}}_{a+da}-{\cal{P}}_a$ then
PEP can describe a tunneling process too.}. The mechanism for
transition from one state to another is through a small external
perturbation\footnote{Certainly, in quantum cosmology, the universe
is considered as one whole \cite{hall1} and the introduction of an
external force is irrelevant. However, because of the lack of a full
theory to describe the universe, these small external forces are
merely used to afford a better understanding of the discussions
presented here.}. To make the discussion above more clear, we take
an example which we shall encounter later on but will present the
result in the form of the following figure. In what follows, we
shall focus on the probabilistic description of our quantum states
provided by figure 1 without worrying about the details of the model
of which the figure is a result.
\begin{figure}[th]
\centerline{\includegraphics[width=10cm]{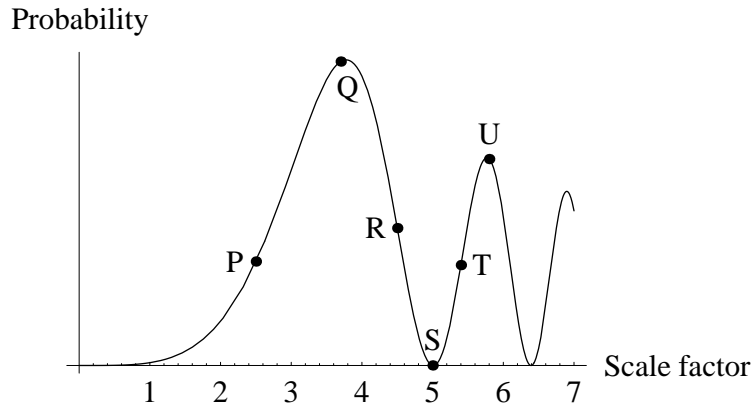}}
\caption{\label{fig1}\footnotesize Figure used to explain the idea
of PEP.}
\end{figure}

Let us take a specific initial condition, say $a=2.5$,
corresponding to the point $P$. Then, PEP states that the system
(here specified by the scale factor $a$) moves continuously to a
state with higher probability and thus $P$ moves to the right to
reach the point  $Q$, a local maximum. Here, since $Q$ locally has
the maximum probability the system stays at $Q$. This means that
the scale factor becomes constant as the time passes\footnote{A
perturbation around the local maximum is acceptable as described
before.}. We show this transition by $P{\buildrel {PEP} \over
\longrightarrow }Q$. Now let the initial condition be the point
$R$. Then we have $R{\buildrel {PEP} \over \longrightarrow }Q$,
and so on. Note that $R{\buildrel {PEP} \over \longrightarrow }S$
is possible but it has much smaller probability compared to the
transitions $R{\buildrel {PEP} \over \longrightarrow }Q$. We note
that the transitions $R{\buildrel {PEP} \over \longrightarrow }S$
and $S{\buildrel {PEP} \over \longrightarrow }T$ may be
interpreted as tunnelling precesses in ordinary quantum mechanics
in the sense that the probability of being at $S$ is zero. It
means that PEP can reproduce tunnelling processes but with a very
small probability.

In the following we will use PEP to describe the evolution of
quantum cosmological states in the canonical method of quantization.
We insist that PEP plays a crucial role in the interpretation of the
states in canonical quantization and allows them to  be compared to
the states resulting from quantization by deformation of the
phase-space structure.
\section{Phase-space deformation: a procedure for quantization}\label{3}
It has long been argued that a deformation in phase-space can be
seen as an alternative path to quantization, based on Wigner
quasi-distribution function and Weyl correspondence between
quantum-mechanical operators in Hilbert space and ordinary c-number
functions in phase-space, see for example \cite{ quantization} and
the references therein. The deformation in the usual phase-space
structure is introduced by Moyal brackets which are based on the
Moyal product \cite{9}. However, to introduce such deformations it
is more convenient to work with Poisson brackets rather than Moyal
brackets.

From a cosmological point of view, models are built in a
minisuper-(phase)-space. It is therefore safe to say that studying
such a space in the presence of deformations mentioned above can
be interpreted as studying the quantum effects on cosmological
solutions. One should note that in gravity the effects of
quantization are woven into the existence of a fundamental length
\cite{10}. The question then arises as to what form of
deformations in phase-space is appropriate for studying quantum
effects in a cosmological model? Studies in noncommutative
geometry \cite{11} and generalized uncertainty principle (GUP)
\cite{12}  have been a source of inspiration for those who have
been seeking an answer to the above question. More precisely,
introduction of modifications in the structure of geometry in the
way of noncommutativity has become the basis from which similar
modifications in the phase-space have been inspired. In this
approach, the fields and their conjugate momenta play the role of
coordinate basis in noncommutative geometry \cite{13}. In doing so
an effective model is constructed whose validity will depend on
its power of prediction. For example, if in a model field theory
the fields are taken as noncommutative, as has been done in
\cite{13}, the resulting effective theory predicts the same
Lorentz violation as a field theory in which the coordinates are
considered as noncommutative \cite{14}. As a further example, it
is well known that string theory can be used to suggest a
modification in the bracket structure of coordinates, also known
as GUP \cite{12} which is used to modify the phase-space structure
\cite{15}. Over the years, a large number of works on
noncommutative fields \cite{9} have been inspired by
noncommutative geometry model theories \cite{11}.

In this paper we study the effects of the existence of a
fundamental length in a cosmological scenario by constructing a
model based on the noncommutative structure of the Deformed
(Doubly) Special Relativity \cite{dsr} which is related to what is
known as the $\kappa$-deformation \cite{16}. This way of
introducing noncommutativity is interesting because of its
compatibility with Lorentz symmetry, as is commonly believed
\cite{16,17}. The $\kappa$-deformation is introduced and studied
in \cite{18}. The $\kappa$-Minkowski space \cite{19} arises
naturally from the $\kappa$-Poincare algebra \cite{dsr} such that
the ordinary Poisson brackets between the coordinates are replaced
by
\begin{equation}\label{1}
\left\{x_0,x_i\right\}=\frac{1}{\kappa}x_i,
\end{equation}
where $\kappa$ is the deformation (noncommutative) parameter which
has the dimension of mass $\kappa=\epsilon \ell^{-1}$ when
$c=\hbar=1$, and $\epsilon=\pm 1$ \cite{20} such that $\kappa$ and
$\ell$ are interpreted as dimensional parameters which are
identified with the fundamental energy and length, respectively. As
mentioned above, one can change the structure of the phase-space
based on equation (\ref{1}). Here we will examine a new kind of
modification in the phase-space structure inspired by relation
(\ref{1}), much the same as has been done in \cite{9,15,nima}. In
what follows we introduce noncommutativity based on
$\kappa$-Minkowskian space and study its consequences on the de
Sitter and dusty FRW cosmologies.

Let us start by briefly studying the ordinary, spatially flat FRW
model where the metric is given by
\begin{eqnarray}\label{de Sitter}
ds^2=-N^2(t)dt^2+a^2(t)(dx^2+dy^2+dz^2),
\end{eqnarray}
with $N(t)$ being the lapse function. The Einstein-Hilbert
Lagrangian with a general energy density $V(a)$ becomes
\begin{eqnarray}\label{lagrangian}
{\cal{L}}&=&\sqrt{-g}(R[g]-V(a))\nonumber\\
&=&-6N^{-1}a\dot{a}^2- Na^3 V(a),
\end{eqnarray}
where $R[g]$ is the Ricci scalar and in the second line the total
derivative term has been ignored. The corresponding Hamiltonian up
to a sign becomes
\begin{eqnarray}\label{hamiltonian1}
{\cal{H}}_0&=&\frac{1}{24}Na^{-1}p_a^2- Na^3V(a).
\end{eqnarray}
Here, we note that since the momentum conjugate to $N(t)$,
$\pi=\frac{\partial{\cal{L}}}{\partial \dot{N}}$ vanishes, the
term $\lambda \pi$ must be added as a constraint to Hamiltonian
(\ref{hamiltonian1}). The Dirac Hamiltonian then becomes
\begin{eqnarray}\label{hamiltonian}
{\cal{H}}&=&\frac{1}{24}Na^{-1}p_a^2- Na^3V(a)+\lambda \pi.
\end{eqnarray}
To introduce noncommutativity one can start with
\begin{eqnarray}\label{kappa-nc-filds}
\{N'(t),a'(t)\}= \ell a'(t).
\end{eqnarray}
This is similar to equation (\ref{1}) since one can interpret $N(t)$ and $a(t)$, appearing as the coefficients of
$dt$ and $d\vec{x}$, in the same manner as $x_0$ and $x_i$
appearing in (\ref{1}) respectively. For this reason we shall call
it the $\kappa$-Minkowskian-minisuper-phase-space. In this case
the Hamiltonian becomes
\begin{eqnarray}\label{primedhamiltonian}
{\cal{H}}'_0&=&\frac{1}{24}N'a'^{-1}{p'}_{a}^2- N'{a'}^3 V(a'),
\end{eqnarray}
where the ordinary Poisson brackets are satisfied except in
(\ref{kappa-nc-filds}). To move along, one introduces the
following variables \cite{jcap,pla,21}
\begin{eqnarray}\label{newvariables}
\left\{
\begin{array}{ll}
N'(t)=N(t)-\ell a(t) p_a(t),\\
a'(t)=a(t),\\
p'_a(t)=p_a(t).
\end{array}\right.
\end{eqnarray}
It can be easily checked that the above variables will satisfy
(\ref{kappa-nc-filds}) if the unprimed variables satisfy the
ordinary Poisson brackets. The term  $-\ell a(t) p_a(t)$ may be
looked upon as a direct consequence of a phase-space deformation
of relation (\ref{kappa-nc-filds}) which, as has been suggested,
could originate from string theory, noncommutative geometry and so
on, see \cite{quantization,nima,chern}. With the above
transformations, Hamiltonian (\ref{primedhamiltonian}) takes the
form
\begin{eqnarray}\label{hamiltonianNC1}
{\cal{H}}^{nc}_0=\frac{1}{24}Na^{-1}{p}_{a}^2-
N{a}^3V(a)-\frac{1}{24}\ell p_a^3+\ell a^4 V(a) p_a.
\end{eqnarray}
It is clear that the momentum conjugate to $N(t)$ does not appear
in (\ref{hamiltonianNC1}), {\it i.e.} $\pi =0$ is a primary
constraint. It can be checked by using Legendre transformations
that the conjugate momentum corresponding to $N(t)$,
$\pi=\frac{\partial{\cal{L}}}{\partial \dot{N}}$, vanishes. It is
therefore necessary to add the term $\lambda \pi$ to Hamiltonian
(\ref{hamiltonianNC1}) to obtain the Dirac Hamiltonian
\begin{eqnarray}\label{hamiltonianNC}
{\cal{H}}^{nc}=\frac{1}{24}Na^{-1}{p}_{a}^2-
N{a}^3V(a)-\frac{1}{24}\ell p_a^3+\ell a^4 V(a) p_a+\lambda\pi.
\end{eqnarray}
The equations of motion resulting from Hamiltonian
(\ref{hamiltonianNC}) are
\begin{eqnarray}\label{NCeqofmotion}
\dot{a}&=&\left\{a,{\cal{H}}^{nc}\right\}=\frac{1}{12}Na^{-1}p_a-
\frac{1}{8}\ell p_a^2+\ell a^4V(a),\nonumber\\
\dot{p_a}&=&\left\{p_a,{\cal{H}}^{nc}\right\}=\frac{1}{24}Na^{-2}p_a^2+3N
a^2V(a)+Na^3V'(a)-4\ell a^3V(a)p_a-\ell a^4V'(a)p_a,\nonumber\\
\dot{N}&=&\left\{N,{\cal{H}}^{nc}\right\}=\lambda,\nonumber\\
\dot{\pi}&=&\left\{\pi,{\cal{H}}^{nc}\right\}=-\frac{1}{24}a^{-1}p_a^2+a^3V(a),\label{NCconstraint}
\end{eqnarray}
where a prime denotes differentiation with respect to the argument.
The requirement that the primary constraints should hold during the
evolution of the system means that $\dot{\pi}=\left\{\pi,{\cal
H}^{nc}\right\}= 0$. If $p_a$ is now calculated from the secondary
constraint, $\dot{\pi} = 0$, and the result is substituted in the
first equation in (\ref{NCeqofmotion}) one obtains
\begin{eqnarray}\label{equationNC1111}
\dot{a}+2\ell a^4V(a)=N\sqrt{\frac{1}{6}a^2V(a)}.
\end{eqnarray}
It is easy to check that the above equation is consistent with the second equation in (\ref{NCeqofmotion}) as well.

\subsection{The de Sitter model}
\begin{figure}
\begin{tabular}{ccc} \epsfig{figure=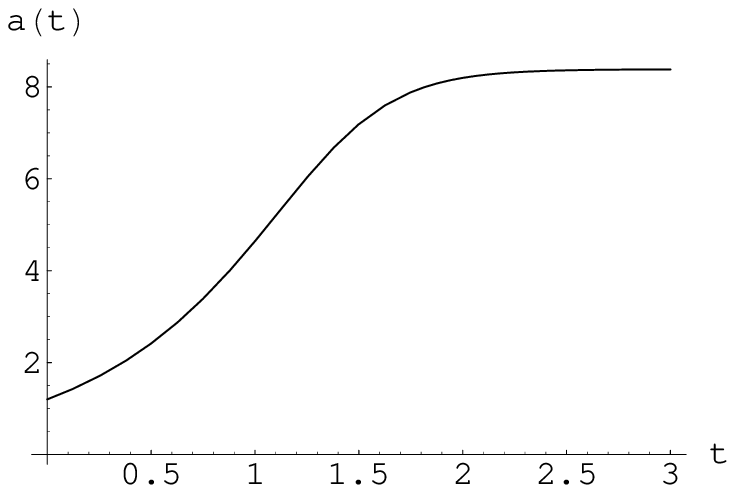,width=7cm}
\hspace{1cm} \epsfig{figure=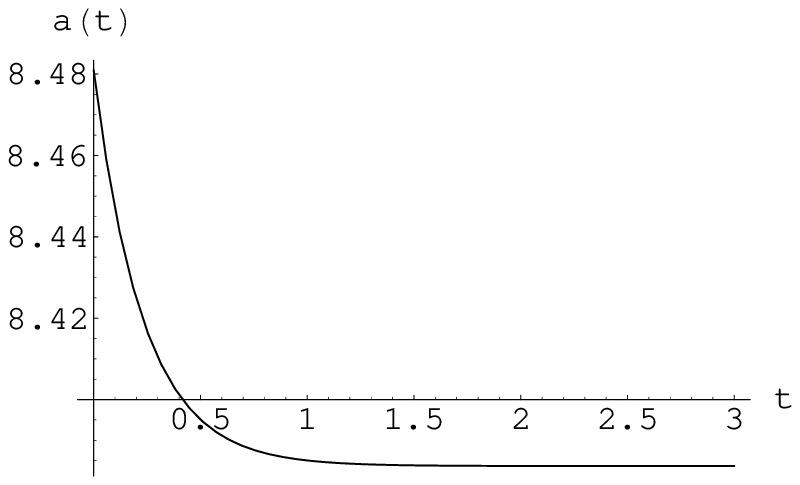,width=7cm}
\end{tabular}
\caption{\label{lambda1}\footnotesize The scale factor for
$\ell=0.0001$ in the de Sitter model. The initial conditions are
$C_1=1$ and $\Lambda=6$ for the left graph and $C_1=-0.0001$ for the
right one.}
\end{figure}
\begin{figure}
\begin{tabular}{ccc} \epsfig{figure=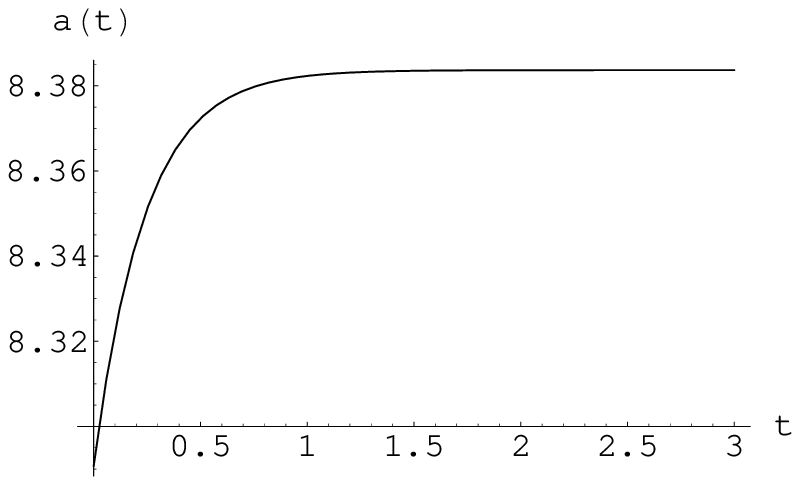,width=7cm}
\hspace{1cm} \epsfig{figure=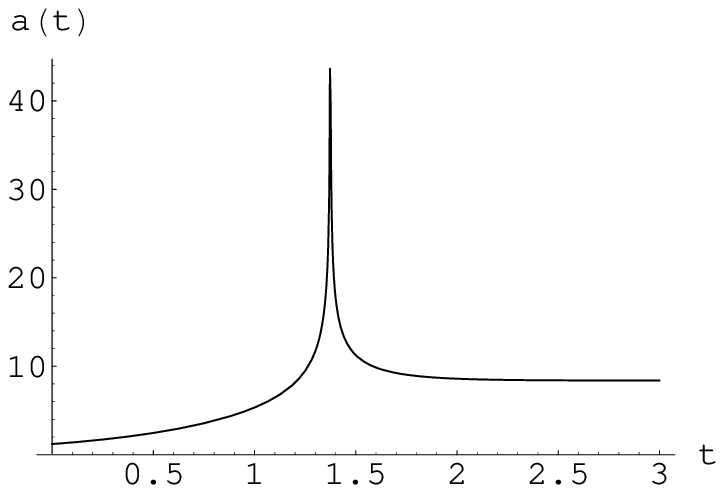,width=7cm}
\end{tabular}
\caption{\label{lambda3}\footnotesize The scale factor for
$\ell=-0.0001$ in the de Sitter model. The initial conditions are
$C_1=-0.0001$ and $\Lambda=6$ for the left graph and $C_1=1$ for the
right one.}
\end{figure}
As is well known, in the de Sitter model $V(a)=2\Lambda$ and
equation (\ref{equationNC1111}) becomes
\begin{eqnarray}\label{equationNC}
\dot{a}+4\ell \Lambda a^4=N\sqrt{\frac{1}{3}\Lambda a^2},
\end{eqnarray}
which is compatible with other equations. The solution of the above
equation, the scale factor, for a constant $N$ can be written
as\footnote{\label{referee}This means that we restrict ourselves to
a certain class of gauges, namely $N=const.$, which is equivalent to
the choice $\lambda=0$ in the equation of motion
(\ref{NCeqofmotion}).}
\begin{eqnarray}\label{k=0}
a(t)=3^{\frac{1}{6}}\left(\frac{N e^{\sqrt{3\Lambda}\hspace{.5mm}N
t} }{C_1+12\ell \sqrt{\Lambda} e^{\sqrt{3\Lambda}\hspace{.5mm}N t}
}\right)^{\frac{1}{3}},
\end{eqnarray}
where $C_1$ is a constant of integration. It is seen that the usual
de Sitter solution (without any deformations) is recovered in the
limit $\ell\rightarrow0$. The scale factor calculated in (\ref{k=0})
and its behavior are shown in figure \ref{lambda1} for $\ell>0$ and
in figure \ref{lambda3} for $\ell<0$. In what follows, we discuss
the different figures separately. This may seem irrelevant here
since the behavior of the scale factors for different choices are
similar to each other. However, in the next case when we study the
dust model, the solutions are essentially different, demanding
separate discussions on their behavior. The same is also true when
we study the relation of the deformed phase-space solutions to that
of the canonical method at the end of next section.

The left plot in figure \ref{lambda1} shows that the scale factor
has started from an initial value and becomes constant in the end.
During this passage the behavior of the scale factor is
monotonically increasing. The right plot in figure \ref{lambda1} on
the other hand, shows a conversing behavior and predicts a
monotonically decreasing scale factor. Note that in this case the
scale factor becomes constant in its final state too. The scale
factor represented by the left plot in figure \ref{lambda3} has the
same behavior as that in the left plot in figure \ref{lambda1}. In
figure \ref{lambda3} one must restrict the discussion  to the second
part of the right plot only since the denominator of the scale
factor in (\ref{k=0}) cannot be zero for physical reasons. This
means, for example, that the universe has come into being at $t=1.5$
in our case. Now the behavior of the scale factor is the same as
that of the scale factor in the right plot of figure \ref{lambda1}.
Note that in all the figures we have taken $N=1$ which only  changes
the values of the initial and final scale factor without any change
in the behavior\footnote{This is compatible with the spirit of gauge
invariance, that is, there is no physical difference between
multiple gauge fixing. The measured value of the system parameters
could be different for different gauges but as mentioned before,
they have the same behavior.}. It should be mentioned here that
these different initial and final values are crucial to our
discussion on the equivalence between these two different
approaches.
\subsection{The dust model}
\begin{figure}
\begin{tabular}{ccc} \epsfig{figure=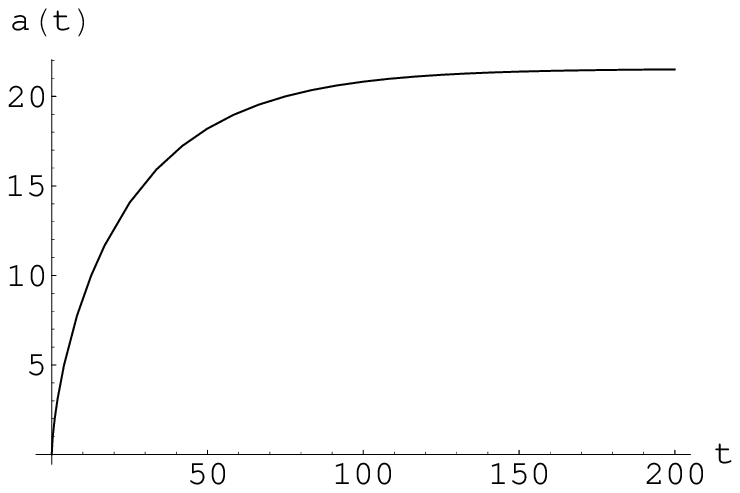,width=7cm}
\hspace{1cm} \epsfig{figure=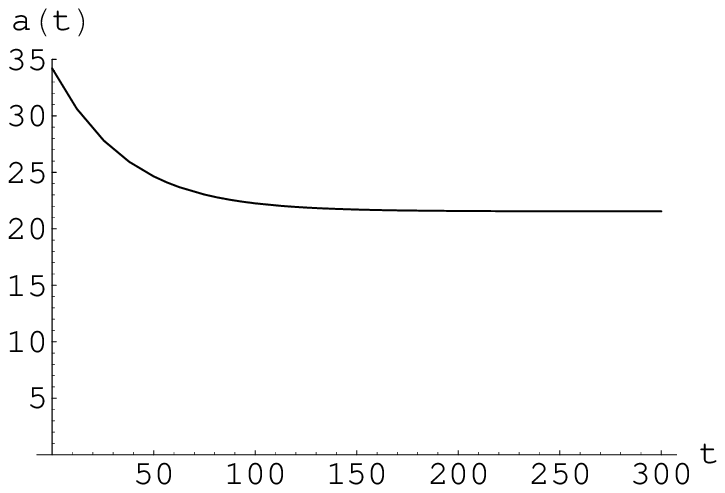,width=7cm}
\end{tabular}
\caption{\label{frw1}\footnotesize The scale factor for $\ell=0.01$
in the dust model. The initial conditions are $C_2=1$ and $\rho_0=1$
for the left graph and $C_2=-1$ for the right one.}
\end{figure}

\begin{figure}[th]
\centerline{\includegraphics[width=7cm]{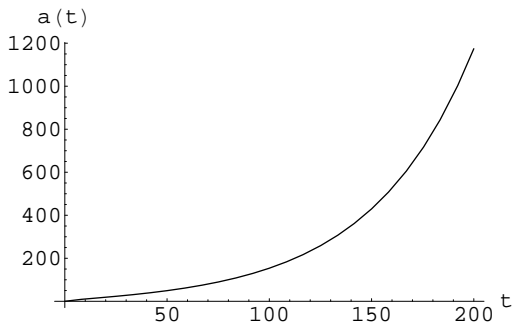}}
\caption{\label{frw0}\footnotesize The scale factor for
$\ell=-0.01$ in the dust model. The initial conditions are $C_2=1$
and $\Lambda=6$. Note that in this case, for all the valid values
of $C_2$ ($C_2\geq1$) we have the same behaving graphs.}
\end{figure}

In this case, substituting $V(a)=\rho_0a^{-3}$ in equation
(\ref{equationNC1111}) results in
\begin{eqnarray}\label{equationNC1}
\dot{a}+2\ell \rho_0 a=N\sqrt{\frac{\rho_0}{6a}}.
\end{eqnarray}
The solution for $N=const.$ is given by
\begin{eqnarray}\label{nc.solutions1}
a(t)=\frac{N^{\frac{2}{3}}}{2\times3^{\frac{1}{3}}}\left[\frac{1-
{C_2}
e^{-3\ell\rho_0\vspace{2mm}t}}{\ell\sqrt{\rho_0}}\right]^{\frac{2}{3}},
\end{eqnarray}
for $\ell>0$ and
\begin{eqnarray}\label{nc.solutions11}
a(t)=\frac{N^{\frac{2}{3}}}{2\times3^{\frac{1}{3}}}\left[\frac{-1+
{C_2}
e^{3|\ell|\rho_0\vspace{2mm}t}}{|\ell|\sqrt{\rho_0}}\right]^{\frac{2}{3}},
\end{eqnarray}
for $\ell<0$, where $C_2$ is an integration constant which, for the
second solution (\ref{nc.solutions11}), must satisfy
${C_2}e^{3|\ell|\rho_0\vspace{2mm}t}\geq1$, indicating that
$C_2\geq1$. Figures \ref{frw1} and \ref{frw0} show the behavior of
the scale factors for $\ell>0$ and $\ell<0$ respectively. The scale
factor presented by the left plot in figure \ref{frw1} suggests that
the universe starts from an initial state and reaches a constant
final state due to a monotonically increasing behavior. As can be
seen from the right plot in figure \ref{frw1}, the converse is also
true. It is worth to consider that only for $C_2=1$ the limitation
$\ell\rightarrow 0$ results in exactly the same behavior due to its
counterpart in absence of any deformations\footnote{It does not make
other choices for $C_2$ irrelevant since the limiting process is yet
a questionable matter. Specifically, the solution which is
represented in the right figure in figure \ref{frw1}, has not any
correspondents in non-deformed phase-space and it is a prediction of
quantum cosmology only.}.

As mentioned in the previous section, the FRW dust model
represents a completely different behavior for $\ell<0$, which is
represented in figure \ref{frw0}. In this case the scale factor
monotonically increases without reaching a final constant value.

\subsection{The radiation model}
\begin{figure}
\begin{tabular}{ccc} \epsfig{figure=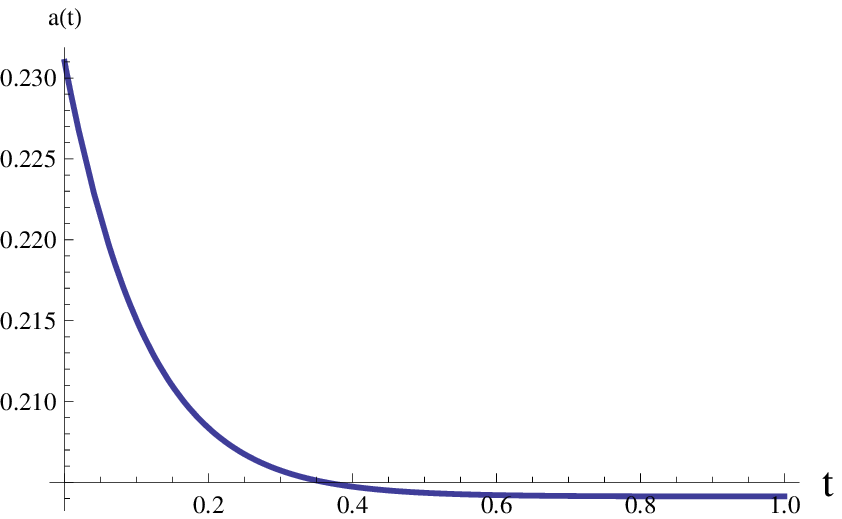,width=7cm}
\hspace{1cm} \epsfig{figure=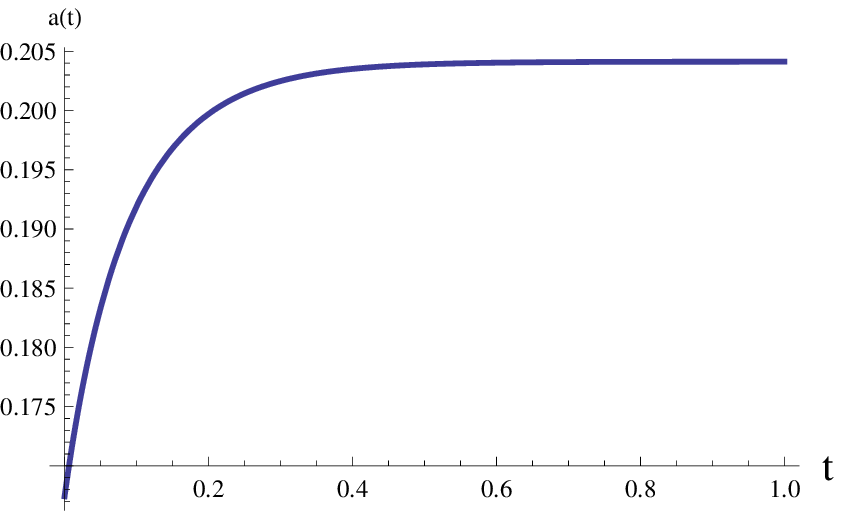,width=7cm}
\end{tabular}
\caption{\label{rad1}\footnotesize The scale factor for $\ell=1$,
$\rho_0=1$ and $C_3=1$ in the radiation model is shown in the left
figure. In the right figure the scale factor for $C_3=-1$ is shown.}
\end{figure}

\begin{figure}
\begin{tabular}{ccc} \epsfig{figure=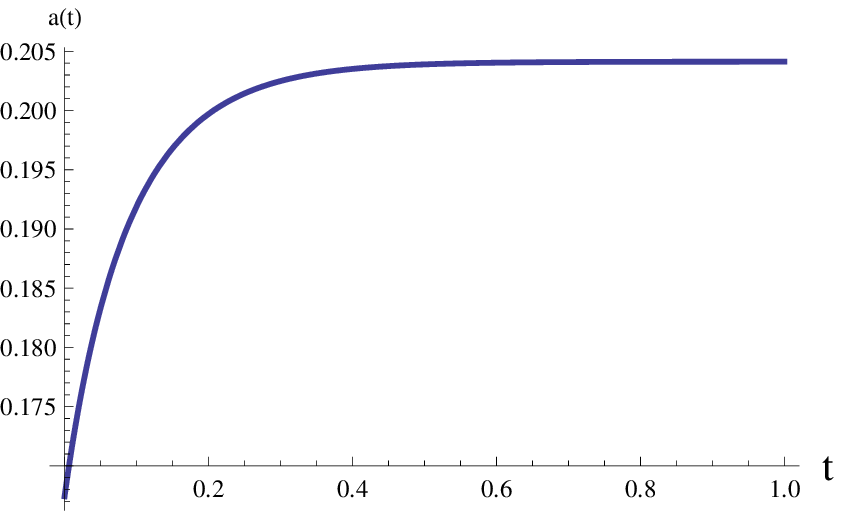,width=7cm}
\hspace{1cm} \epsfig{figure=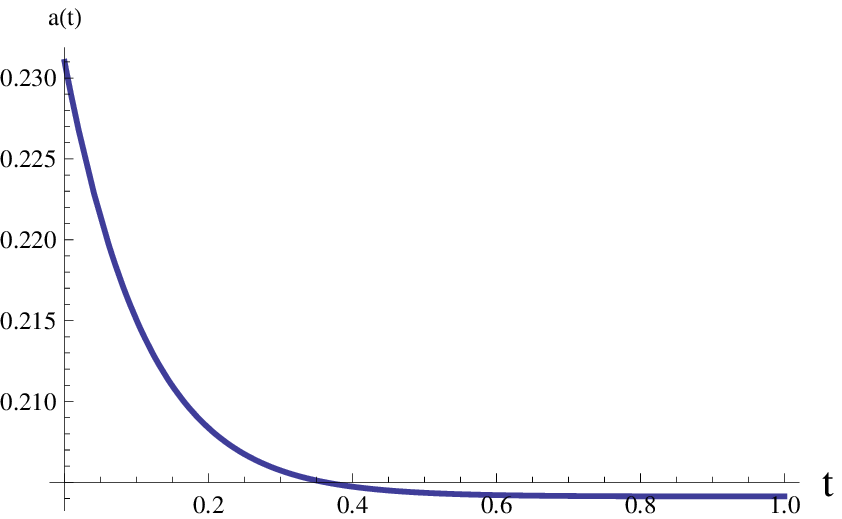,width=7cm}
\end{tabular}
\caption{\label{rad2}\footnotesize The scale factor for $\ell=-1$,
$\rho_0=1$ and $C_3=1$ in the radiation model is shown in the left
figure. In the right figure the scale factor for $C_3=1$ is shown.
Note that since in the metric (\ref{de Sitter}), $a^2(t)$ is
appeared so the absolute value of the scale factor
(\ref{nc.solution32}) is meaningful.}
\end{figure}

This case is more interesting since it aims to describe the early
phases of the Universe, when quantum geometrical properties are
expected to dominate. Mathematically, for radiation era $V(a)=\rho_0
a^{-4}$ which makes equation (\ref{equationNC1111})
\begin{eqnarray}\label{equationNC31}
\dot{a}+2\ell \rho_0 =\sqrt{\frac{\rho_0}{6}} N a^{-1}.
\end{eqnarray}
The solution of the above equation  reads as
\begin{eqnarray}\label{nc.solution31}
a(t)=\frac{\sqrt{6} N}{12 \ell
\sqrt{\rho_0}}\left[1+{\cal{W}}\left(\frac{C_3 e^{(-1-4\sqrt{6}
N^{-1} \ell^2\rho_0^{\frac{3}{2}}t)}}{\sqrt{6}N}\right)\right],
\end{eqnarray}
for $\ell>0$ and it reads
\begin{eqnarray}\label{nc.solution32}
a(t)=\frac{\sqrt{6} N}{12 \ell
\sqrt{\rho_0}}\left[1+{\cal{W}}\left(-\frac{C_3 e^{(-1-4\sqrt{6}
N^{-1} \ell^2\rho_0^{\frac{3}{2}}t)}}{\sqrt{6}N}\right)\right],
\end{eqnarray}
and for $\ell<0$ where $C_3$ is an integration constant,
${\cal{W}}(z)$ is Lambert W function and gives the principal
solution for $w$ in $z=we^w$. Figures \ref{rad1} and \ref{rad2} show
the behavior of scale factor for $\ell>0$ and $\ell<0$,
respectively.

\section{Canonical quantization: the WD equation}\label{ca}
We now focus attention on the study of the quantum cosmology of the
models described above. For this purpose we quantize the dynamical
variables with the use of the WD equation, that is, ${\cal
H}\Psi=0$, where ${\cal H}$ is the operator form of the Hamiltonian
given by equation (\ref{hamiltonian1}), and $\Psi$ is the wave
function of the universe, a function of the scale factor and the
matter fields, if they exist.
\subsection{The de Sitter model}\label{wer}
\begin{figure}
\begin{tabular}{ccc} \epsfig{figure=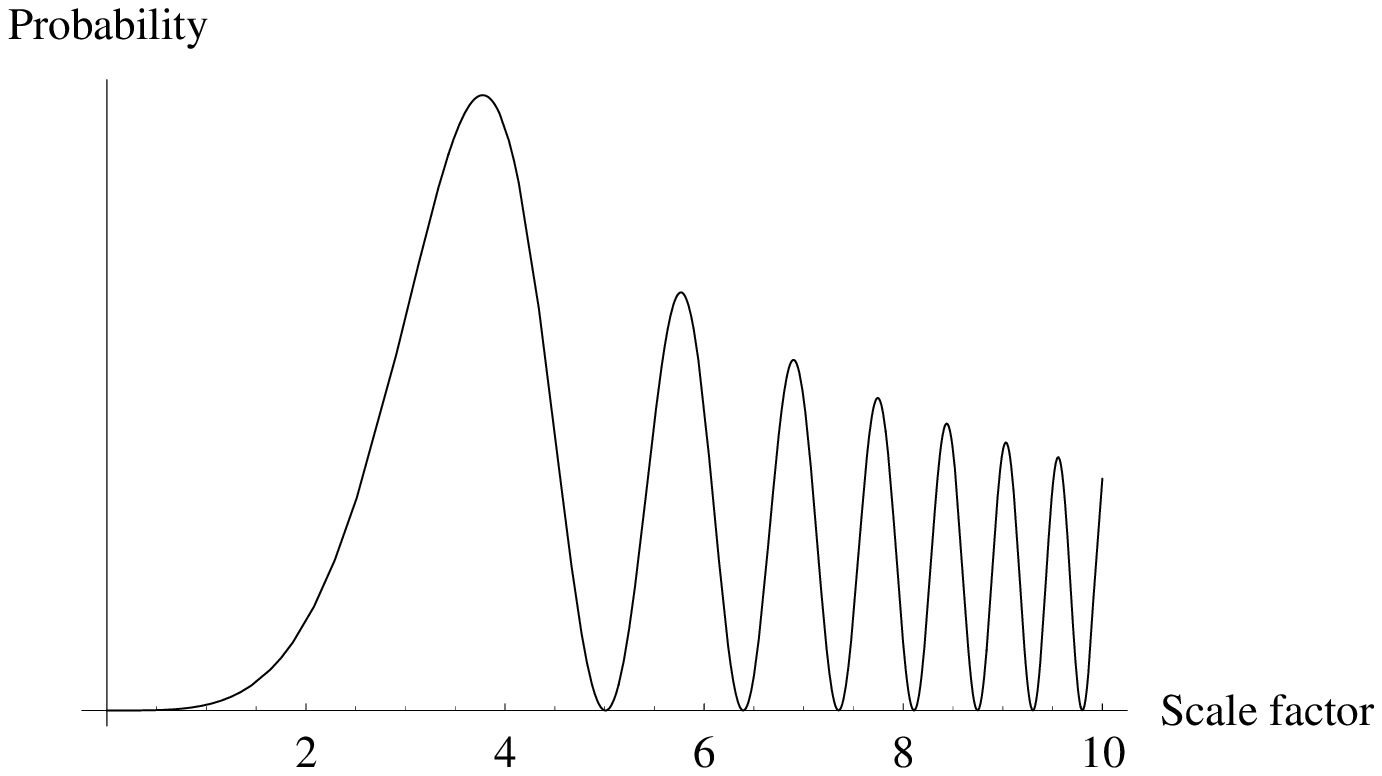,width=7cm}
\hspace{1cm} \epsfig{figure=fig1.eps,width=7cm}
\end{tabular}
\caption{\label{J+}\footnotesize The probability density for the
$J_{\frac{1}{3}}$ term with $\Lambda=0.0001$.}
\end{figure}

\begin{figure}
\begin{tabular}{ccc} \epsfig{figure=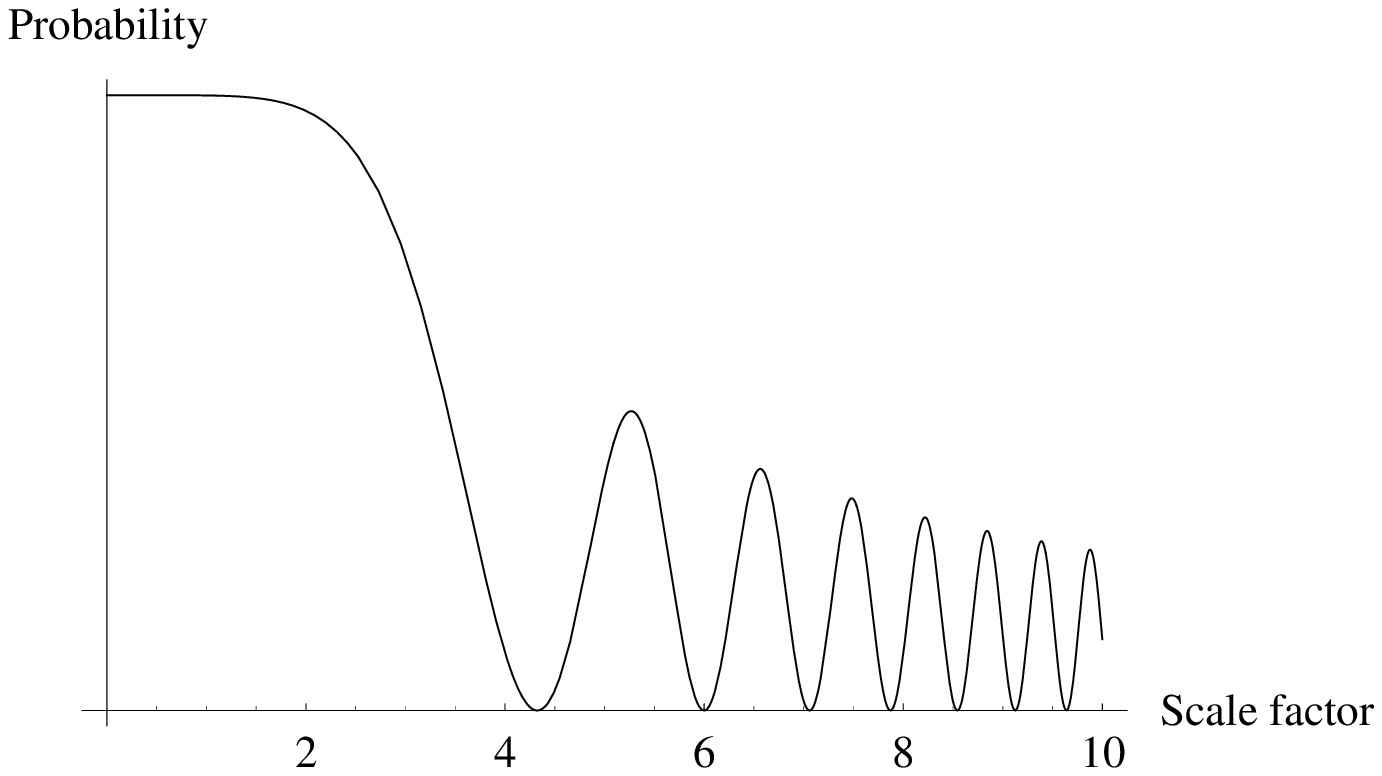,width=7cm}
\hspace{1cm} \epsfig{figure=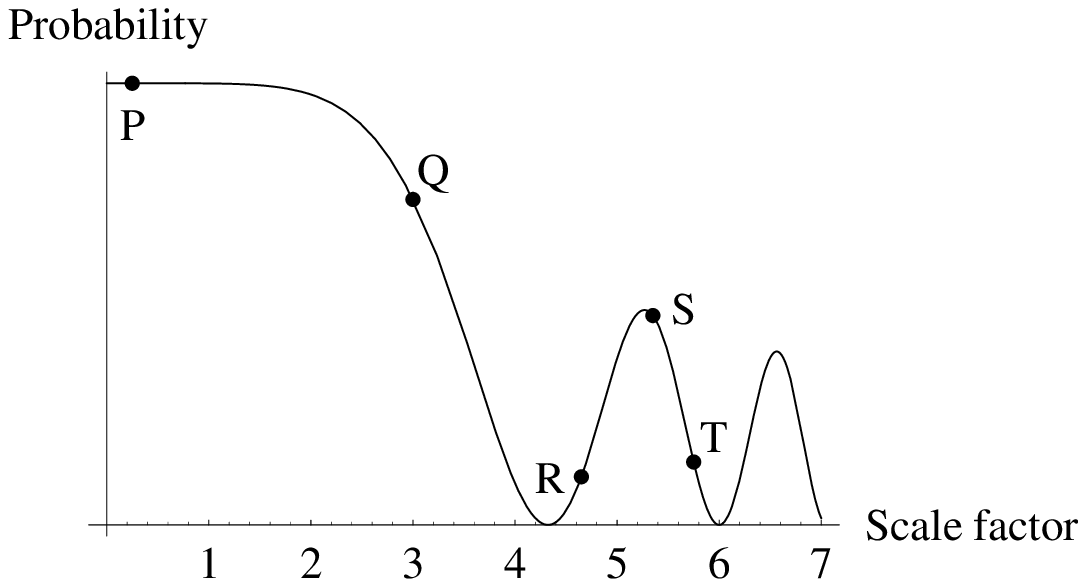,width=7cm}
\end{tabular}
\caption{\label{J-}\footnotesize The probability density for the
$J_{-\frac{1}{3}}$ term with $\Lambda=0.0001$.}
\end{figure}

Let us set $V(a)=2\Lambda$. Then the corresponding WD equation
becomes
\begin{eqnarray}\label{WD}
{\cal{H}}\Psi(a)=\left[\frac{1}{24}a^{-1}p_a^2- 2\Lambda
a^3\right]\Psi(a)=0.
\end{eqnarray}
Choice of the ordering $a^{-1}p_a^2=p_aa^{-1}p_a$ to make the
Hamiltonian Hermitian and use of $[a,p_a]=i\hbar$ and
$p_a=-i\hbar\partial_a$ results in
\begin{eqnarray}\label{diffWD}
\partial_a^2\Psi(a)-a^{-1}\partial_a\Psi(a)+\frac{48\Lambda}{\hbar^2} a^4 \Psi(a)=0,
\end{eqnarray}
with solutions
\begin{eqnarray}\label{solutiondiffWD}
\Psi(a)=c_1a
J_{\frac{1}{3}}\left(\frac{4a^3\sqrt{\Lambda}}{\hbar\sqrt{3}}\right)+
c_2aJ_{-\frac{1}{3}}\left(\frac{4a^3\sqrt{\Lambda}}{\hbar\sqrt{3}}\right),
\end{eqnarray}
where $J$ is the Bessel function and $c_1$ and $c_2$ are arbitrary
constants. The constants must be so chosen as to satisfy the
initial conditions. However, in what follows, our discussions are
independent of the values of these constants. Here we show and
emphasize the different behavior of both solutions. To do this we
need to calculate the probability density $|\Psi(a)|^2$. In
figures \ref{J+} and \ref{J-} the probability density for $c_2=0$
and $c_1=0$ are shown respectively.

Now, we start form a chosen initial condition, say, $a=1$. The PEP
procedure may now be employed to describe the physical
interpretation of the resulting states. In figure \ref{J+}, let the
initial state be at point $P$. One then expects, keeping in mind the
discussion presented previously, the transition $P{\buildrel {PEP}
\over \longrightarrow }Q$ to appear, with the following physical
interpretation. The point $P$ shows an initial state and the point
$Q$ shows the final state. The universe begins its evolution by a
monotonically increasing scale factor to reach the point $Q$. Point
$Q$ is a state with a greater scale factor and is finite. Note that
since the point $Q$ is a local maximum, PEP predicts $Q{\buildrel
{PEP} \over \longrightarrow }Q$ which means that the scale factor
remains in this final state. The transition $P{\buildrel {PEP} \over
\longrightarrow }Q$ then means starting from an initial state,
increasing monotonically, and finally arriving at a constant final
state. This behavior is completely similar to the behavior of the
scale factor in the previous section, presented in the left plot in
figure \ref{lambda1}.

Since for the other cases the details are similar, in the following
we mention only the correspondence between the solutions without any
detail. In figure \ref{J+}, the transition $R{\buildrel {PEP} \over
\longrightarrow }Q$ is similar to the right plot in figure
\ref{lambda1}. On the other hand in the same figure, transition
$T{\buildrel {PEP} \over \longrightarrow }U$ is similar to that of
the left plot in figure \ref{lambda1} with a different initial
condition, that is, with an appropriate $C_1$. The transition
$Q{\buildrel {PEP} \over \longrightarrow }Q$ represents the scale
factor (\ref{k=0}) with $C_1=0$. In figure \ref{J-}, similar
discussions as above apply and the transitions $Q{\buildrel {PEP}
\over \longrightarrow }P$ and $T{\buildrel {PEP} \over
\longrightarrow }S$ have the behavior and interpretation compatible
with the right plot in figure \ref{lambda3} with different initial
conditions. Again, $R{\buildrel {PEP} \over \longrightarrow }S$ is
similar to the left plot in figure \ref{lambda3}.

Now, an interesting question arises as to the meaning of
$P{\buildrel {PEP} \over \longrightarrow }P$ in this case. One
possible answer would be that only superpositions of the solutions
are appropriate for comparison with the previous section results,
e.g. $c_1=c_2=1$ and $c_1=-c_2=1$.

It is appropriate at this point to discuss different values of the
final scale factor predicted by the WD equation. Obviously, since
the maximum probabilities in figures \ref{J+} and \ref{J-} occur at
different values then the final scale factor assumes different
values too. This prediction is compatible with the deformed
phase-space method due to the appearance of the lapse function, $N$,
as a constant parameter in relation (\ref{k=0}) controlling the
behavior of the scale factor. It is worth mentioning that in both
approaches there are two parameters that control the behavior of the
scale factor; $C_1$ in the deformed phase-space method and also $N$
which is fixed by the maximum value of the scale factor and its
initial values in each region of the WD solution.  Note that each
region is an interval between two minima that contains one maximum.
For example in figure 6 the region containing points $P$ and $Q$ is
one region and that containing $T$ and $U$ is another region and the
same is true for other figures.
\subsection{The dust model}
\begin{figure}
\begin{tabular}{ccc} \epsfig{figure=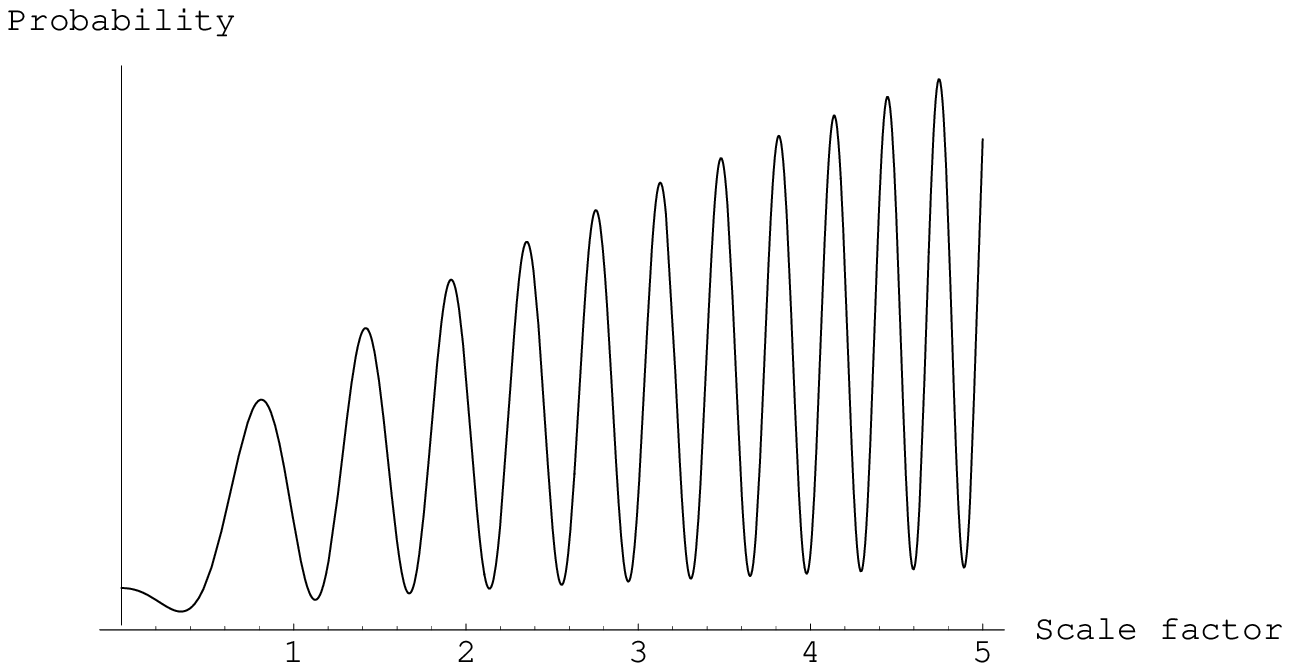,width=7cm}
\hspace{1cm} \epsfig{figure=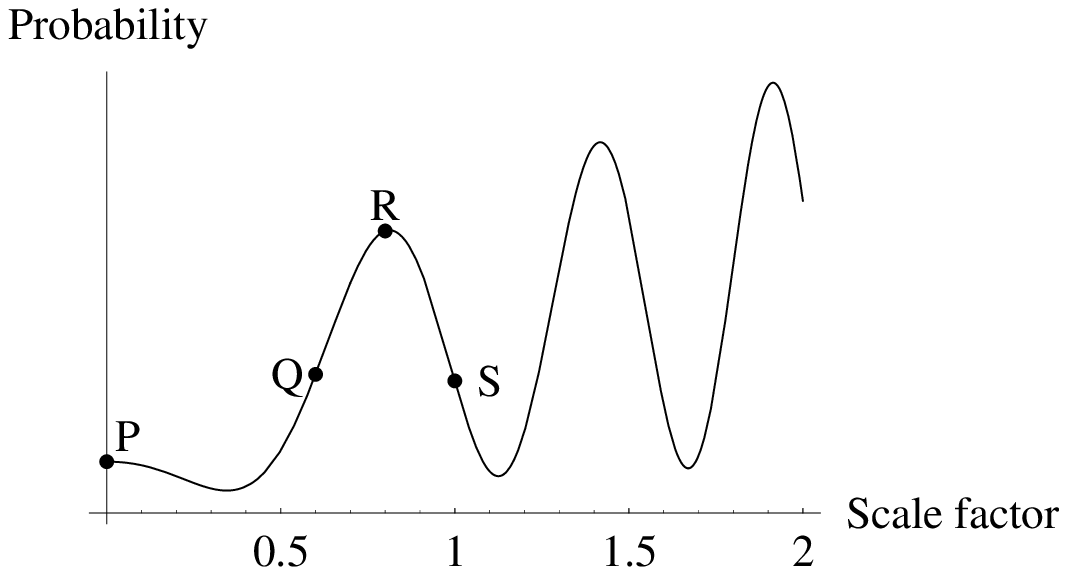,width=7cm}
\end{tabular}
\caption{\label{a}\footnotesize The probability density for the
$\mbox {Bi}'$ term with $\rho_0=1$.}
\end{figure}

\begin{figure}
\begin{tabular}{ccc} \epsfig{figure=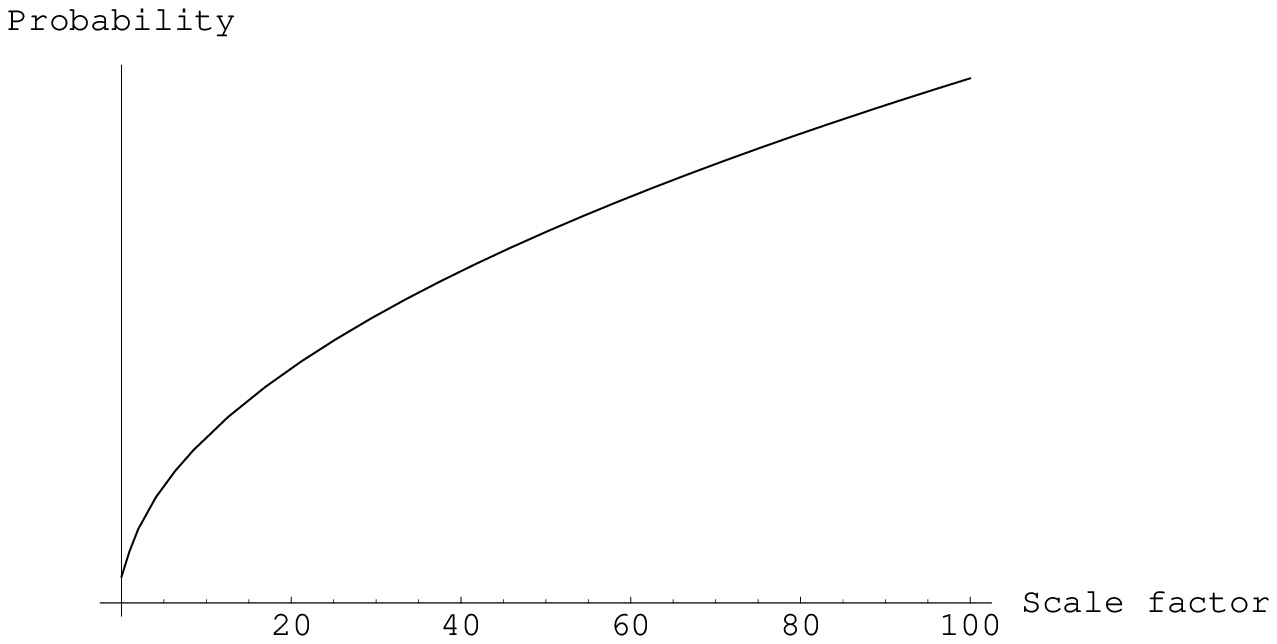,width=7cm}
\hspace{1cm} \epsfig{figure=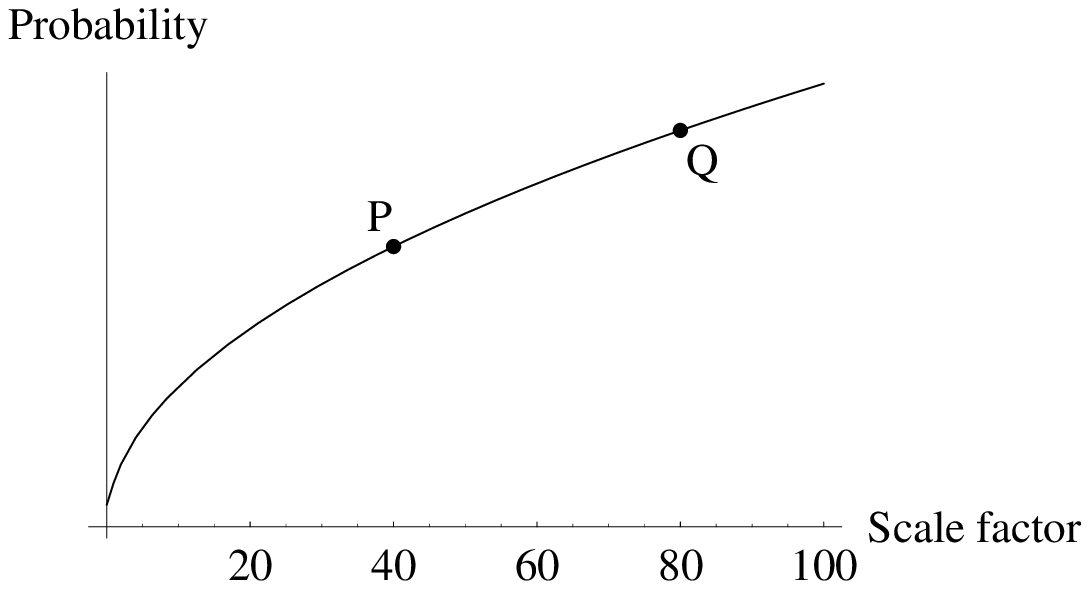,width=7cm}
\end{tabular}
\caption{\label{b}\footnotesize The probability density for the
$\mbox {Ai}'$ term with $\rho_0=1$.}
\end{figure}

In this case we set $V(a)=\rho_0a^{-3}$ in Hamiltonian
(\ref{hamiltonian1}). The corresponding WD equation becomes
\begin{eqnarray}\label{WD1}
{\cal{H}}\Psi(a)=\left[\frac{1}{24}a^{-1}p_a^2-\rho_0\right]\Psi(a)=0.
\end{eqnarray}
With the above ordering, the WD equation is
\begin{eqnarray}\label{diffWD1}
\partial_a^2\Psi(a)-a^{-1}\partial_a\Psi(a)+\frac{24\rho_0}{\hbar^2} a
\Psi(a)=0,
\end{eqnarray}
with solution
\begin{eqnarray}\label{solutiondiffWD1}
\Psi(a)=c'_1 \mbox
{Ai}'\left[2\left(\frac{-3\rho_0}{\hbar^2}\right)^{1/3}a\right]+
c'_2\mbox
{Bi}'\left[2\left(\frac{-3\rho_0}{\hbar^2}\right)^{1/3}a\right],
\end{eqnarray}
where $c'_1$ and $c'_2$ are integration constants and $\mbox
{Ai}'$ and $\mbox {Bi}'$ are derivatives of the Airy functions
${\mbox {Ai}}$ and ${\mbox {Bi}}$ with respect to $a$
respectively. In this case the corresponding probability densities
$|\Psi(a)|^2$ are shown in figures \ref{a} and \ref{b} where
$c'_1=0$ and $c'_2=0$ respectively.
In the dust model two different solutions show completely different
behaviors. One, figure \ref{a}, has local maxima but the other,
figure \ref{b}, is monotonically increasing. In figure \ref{a}, the
transition $Q{\buildrel {PEP} \over \longrightarrow }R$ is
compatible with the left plot in figure \ref{frw1}. Transition
$S{\buildrel {PEP} \over \longrightarrow }R$ however, is compatible
with the right plot in figure \ref{frw1} with transitions
$P{\buildrel {PEP} \over \longrightarrow }P$ and $R{\buildrel {PEP}
\over \longrightarrow }R$ being compatible with scale factor
(\ref{nc.solutions1}) with different initial conditions and
constants.

It is worth noting that the transition $P{\buildrel {PEP} \over
\longrightarrow }Q$ in figure \ref{b} means that no matter what the
initial state is, the scale factor moves to a larger value and
continues to grow since in this case no local maximum exists at all.
This behavior is completely in agreement with the behavior of the
scale factor represented in figure \ref{frw0}. Note that  a similar
discussion like that of the last paragraph of the previous
subsection goes for the dust case. Note that for figure \ref{b}
there is only one region.

\subsection{The radiation model}
\begin{figure}
\begin{tabular}{ccc} \epsfig{figure=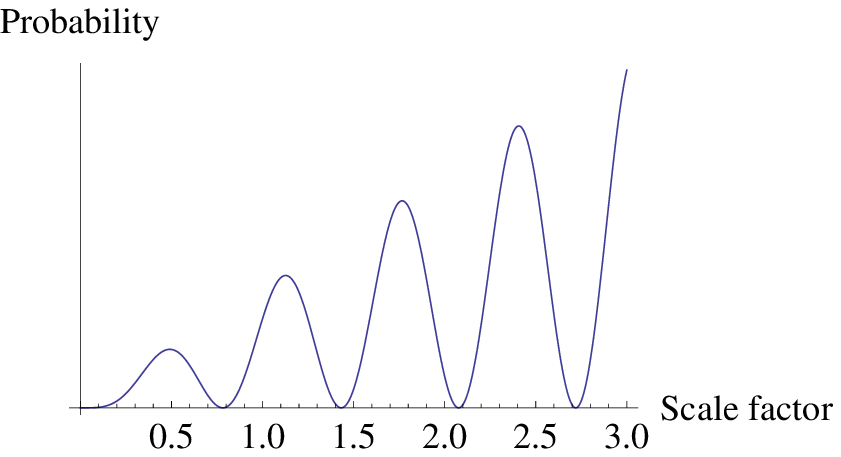,width=7cm}
\hspace{1cm} \epsfig{figure=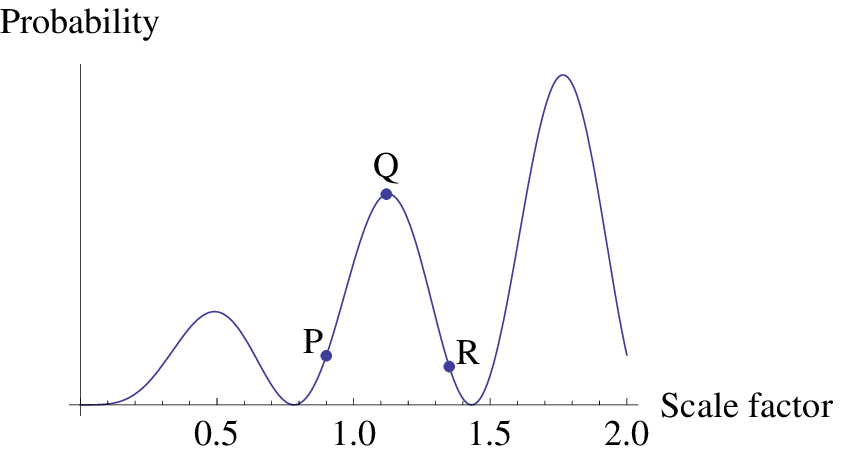,width=7cm}
\end{tabular}
\caption{\label{radq1}\footnotesize The probability density for the
$J$-Bessel term with $\rho_0=1$ and $\hbar=1$.}
\end{figure}

\begin{figure}
\begin{tabular}{ccc} \epsfig{figure=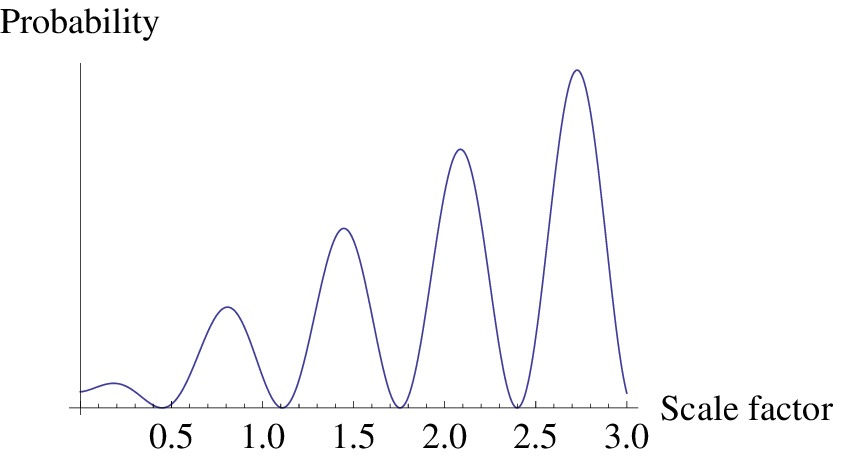,width=7cm}
\hspace{1cm} \epsfig{figure=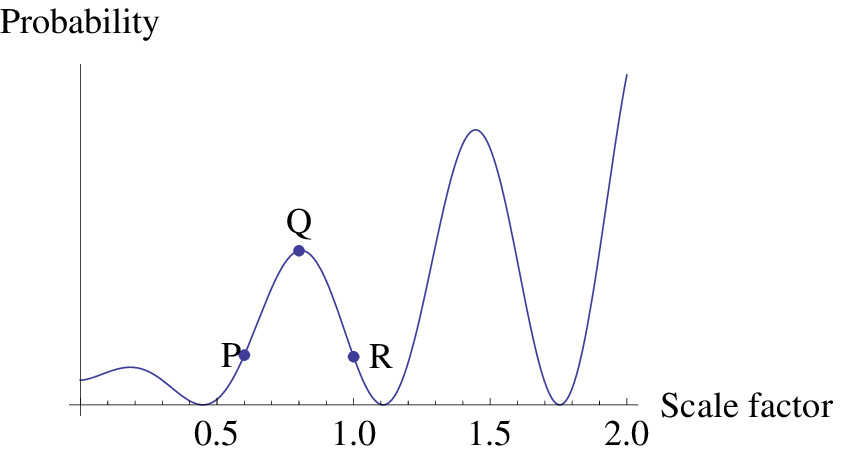,width=7cm}
\end{tabular}
\caption{\label{radq2}\footnotesize The probability density for the
$Y$-Bessel term with $\rho_0=1$ and $\hbar=1$.}
\end{figure}

In the radiation case the Hamiltonian (\ref{hamiltonian1})  with
$V(a)=\rho_0 a^{-4}$ results in the following WD equation
\begin{eqnarray}\label{WD3}
{\cal{H}}\Psi(a)=\left[\frac{1}{24}a^{-1}p_a^2- \rho_0
a^{-1}\right]\Psi(a)=0.
\end{eqnarray}
By using the mentioned ordering it reduces to a differential
equation as follow
\begin{eqnarray}\label{diffWD3}
\partial_a^2\Psi(a)-a^{-1}\partial_a\Psi(a)+\frac{24\rho_0}{\hbar^2} a
\Psi(a)=0,
\end{eqnarray}
with solution
\begin{eqnarray}\label{solutiondiffWD3}
\Psi(a)=c''_1a J_{1}\left(2\sqrt{6 \hbar} a \right)+
c''_2aY_{1}\left(2\sqrt{6 \hbar} a\right),
\end{eqnarray}
where $c''_1$ and $c''_2$ are integration constants and $J$ and $Y$
are Bessel functions. The behavior of the above functions are
represented in figures \ref{radq1} and \ref{radq2}. The discussions
on comparison between quantum cosmological solutions and their
correspondence from deformed phase-space formalism, i.e. figure
\ref{rad1}, are the same as previous models, cosmological constant
and dust models. To prevent making the paper boring we omit
discussing for this case.

\section{Bianchi type I Model}
In this section we consider a more complicated model i.e. Bianchi
model with different matter fields and compare its quantum behavior
that is modeled by a kind of phase-space deformation and also
canonical quantization. We separate Bianchi models from previous
ones since there is more than one variable exists that makes our
conjecture more reasonable.
\subsection{Phase-space deformation}
\begin{figure}
\begin{tabular}{ccc} \epsfig{figure=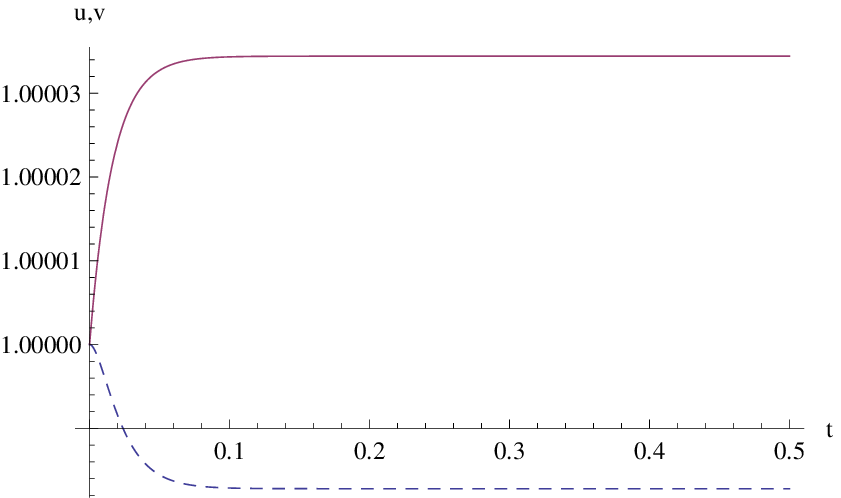,width=7cm}
\hspace{1cm} \epsfig{figure=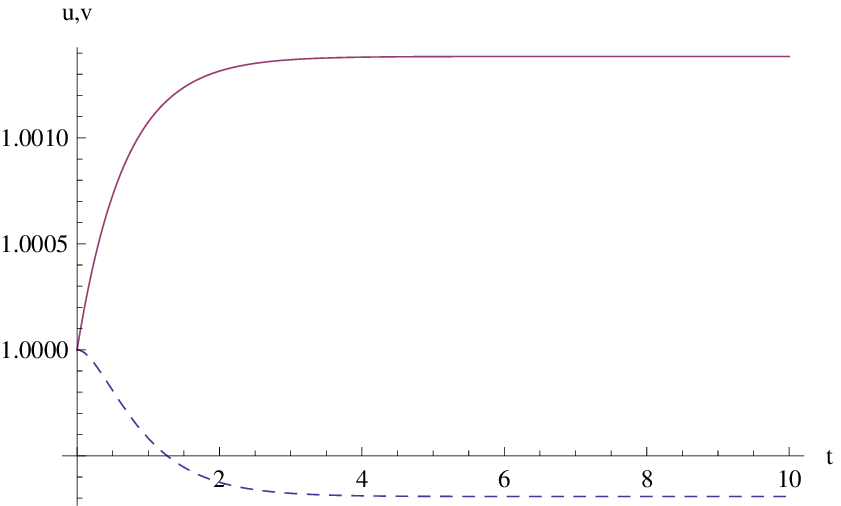,width=7cm}
\end{tabular}
\caption{\label{ch1} \footnotesize Left plot is shown for
$\gamma=-1$ (de Sitter) and right one for $\gamma=0$ (dust). We take
$\alpha=\beta =1/2$, $p_{0v} = 1$, $N=1$, $\ell=1$, $M_{\gamma}=1$
and initial conditions $u(t=0)=v(t=0)=1$, $p_u(t=0)=0$. The dashed
lines represent $u$-behavior and the solid line $v$-behavior in both
figures.}
\end{figure}

\begin{figure}
\begin{tabular}{ccc} \epsfig{figure=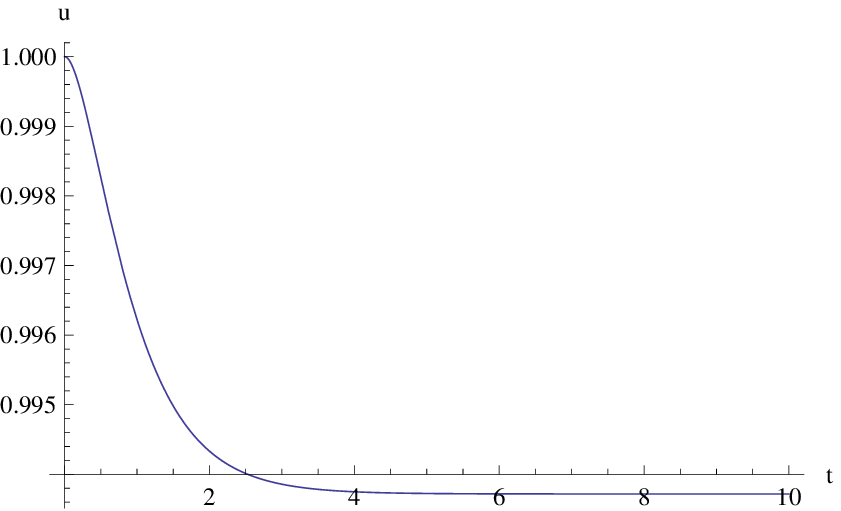,width=7cm}
\hspace{1cm} \epsfig{figure=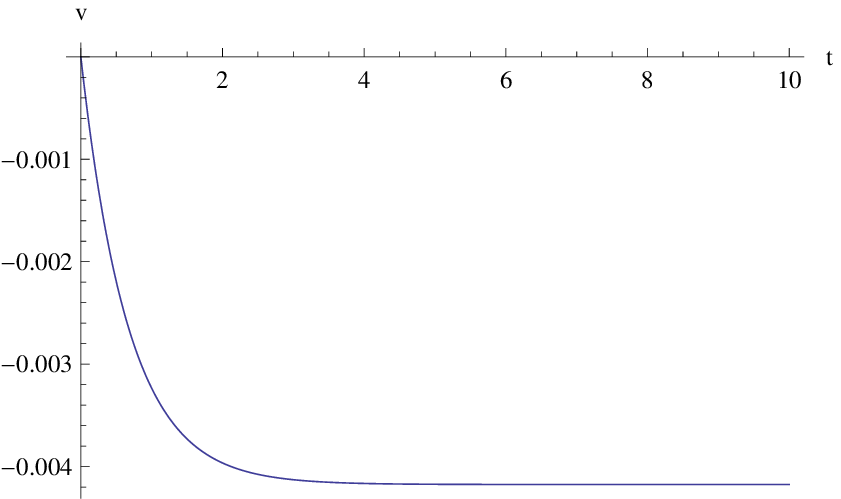,width=7cm}
\end{tabular}
\caption{\label{bianchi2}\footnotesize This figures are sketched
only for dust, $\gamma=0$. We take $\alpha=\beta =1/2$, $p_{0v} =
-1$, $N=1$, $\ell=1$, $M_{\gamma}=1$ and initial conditions
$u(t=0)=v(t=0)=1$, $p_u(t=0)=0$. In the left plot $u$-behavior is
represented and $v$-behavior is represented in the right one.}
\end{figure}

Let us consider a cosmological model in which the spacetime is
assumed to be of Bianchi type I whose metric can be written as
\begin{equation}\label{Bi1}
ds^2=-N^2(t)dt^2+e^{2u(t)}e^{2\beta_{ij}(t)}dx^idx^j,\end{equation}where
$N(t)$ is the lapse function, $e^{u(t)}$ is the scale factor of the
universe and $\beta_{ij}(t)$ determine the anisotropic parameters
$v(t)$ and $w(t)$ as follows
\begin{equation}\label{Bi2}
\beta_{ij}=\mbox{diag}\left(v+\sqrt{3}w,v-\sqrt{3}w,-2v\right).\end{equation}To
simplify the model we take $w=0$, where is equivalent with a
universe with two scale factors in the form
\begin{equation}\label{Bi3}
ds^2=-N^2(t)dt^2+a^2(t)(dx^2+dy^2)+c^2(t)dz^2.\end{equation}The
anisotropy in the above metric can achieved by introducing a large
scale homogeneous magnetic field in a flat FRW spacetime. Such a
magnetic field results in a preferred direction in space along the
direction of the field. If we introduce a magnetic field which has
only a $z$ component, the resulting metric can be written in the
form (\ref{Bi3}) where there are equal scale factors in the
transverse directions $x$ and $y$ and a different one, $c(t)$, in
the longitudinal direction $z$. The Hamiltonian for gravity coupled
to a perfect fluid with equation of state $p=\gamma \rho$ is
\begin{equation}\label{Bi4}
{\cal
H}=\frac{1}{24}Ne^{-3u}\left(-p_u^2+p_v^2\right)+NM_{\gamma}e^{-3\gamma
u}+\lambda \pi,
\end{equation}
where, $M_{\gamma}$ is a model dependent constant. To introduce
noncommutativity one can start with
\begin{equation}\label{Bi5}
\left\{N'(t),e^{u'+v'}\right\}=\ell
e^{u'+v'},\hspace{0.5cm}\left\{N'(t),e^{u'-2v'}\right\}=\ell
e^{u'-2v'},
\end{equation}
where are the two dimensional generalization of relation
(\ref{kappa-nc-filds}). The Hamiltonian of this model becomes
\begin{equation}\label{Bi6}
{\cal H'}_0=\frac{1}{24}N'
e^{-3u'}\left(-{p'}_u^2+{p'}_v^2\right)+N'M_{\gamma}e^{-3\gamma
u'}.\end{equation} Now we introduce as before
\begin{eqnarray}\label{Bi7}
\left\{
\begin{array}{ll}
N'(t)=N(t)-\alpha \ell p_u(t)-\beta \ell p_v(t),\\
u'(t)=u(t),\hspace{0.5cm}v'(t)=v(t),\\
p'_u(t)=p_u(t),\hspace{0.3cm}p'_v(t)=p_v(t)
\end{array}\right.
\end{eqnarray}
where $\alpha+\beta=1$. The deformed Hamiltonian by imposing the
above transformations takes the form
\begin{equation}\label{Bi8}
{\cal H}^{nc}=\left[N(t)-\alpha \ell p_u(t)-\beta \ell
p_v(t)\right]\left[\frac{1}{24}e^{-3u}\left(-p_u^2+p_v^2\right)+M_{\gamma}e^{-3\gamma
u}\right]+\lambda \pi.\end{equation}The classical dynamics is
governed by the Hamiltonian equations, that is
\begin{eqnarray}\label{Bi9}
\left\{
\begin{array}{ll}
\dot{u}=\{u,{\cal H}^{nc}\}=-\frac{1}{12}e^{-3u}p_u\left[N(t)-\alpha \ell p_u(t)-\beta \ell p_v(t)\right]-\alpha \ell \left[\frac{1}{24}e^{-3u}\left(-p_u^2+p_v^2\right)+M_{\gamma}e^{-3\gamma u}\right],\\\\
\dot{v}=\{v,{\cal H}^{nc}\}=\frac{1}{12}e^{-3u}p_v\left[N(t)-\alpha \ell p_u(t)-\beta \ell p_v(t)\right]-\beta \ell \left[\frac{1}{24}e^{-3u}\left(-p_u^2+p_v^2\right)+M_{\gamma}e^{-3\gamma u}\right],\\\\
\dot{p_u}=\{p_u,{\cal H}^{nc}\}=\left[N(t)-\alpha \ell p_u(t)-\beta \ell p_v(t)\right]\left[-\frac{1}{8}e^{-3u}\left(-p_u^2+p_v^2\right)-3\gamma M_{\gamma}e^{-3\gamma u}\right],\\\\
\dot{p_v}=\{p_v,{\cal H}^{nc}\}=0,\\\\
\dot{N}=\{N,{\cal H}^{nc}\}=\lambda,\\\\
\dot{\pi}=\left\{\pi,{\cal{H}}^{nc}\right\}=\frac{1}{24}e^{-3u}\left(-p_u^2+p_v^2\right)+M_{\gamma}e^{-3\gamma
u}.
\end{array}
\right.
\end{eqnarray}
The requirement that the primary constraints should hold during the
evolution of the system means that $\dot{\pi}=\left\{\pi,{\cal
H}^{nc}\right\}= 0$. Applying this condition into the above
equations we get
\begin{eqnarray}\label{Bi10}
\left\{
\begin{array}{ll}
\dot{u}=-\frac{1}{12}e^{-3u}p_u\left[N-\alpha \ell p_u(t)-\beta \ell p_{0v}\right],\\\\
\dot{v}=\frac{1}{12}e^{-3u}p_v\left[N-\alpha \ell p_u(t)-\beta \ell p_{0v}\right],\\\\
\dot{p_u}=-3(\gamma-1)M_{\gamma}e^{-3\gamma u}\left[N-\alpha \ell
p_u(t)-\beta \ell p_{0v}\right].
\end{array}
\right.
\end{eqnarray}
Since the analytic solution does not exist for the above equations,
their behavior is represented in figures \ref{ch1} and
\ref{bianchi2} by applying numerical methods. In figure \ref{ch1}
for $u$, there is a decreasing era before steady behavior in
contrast to $v$-behavior. However, the other choices are possible
too. For example, in figure \ref{bianchi2}, both of $u$ and $v$ are
the same in the existence of a decreasing era in their behaviors.
These figures are plotted for typical numerical values for the
parameters and initial conditions. With examining some other initial
conditions the behavior is almost repeated.

\subsection{Canonical quantization} 
\begin{figure}
\begin{tabular}{ccc} \epsfig{figure=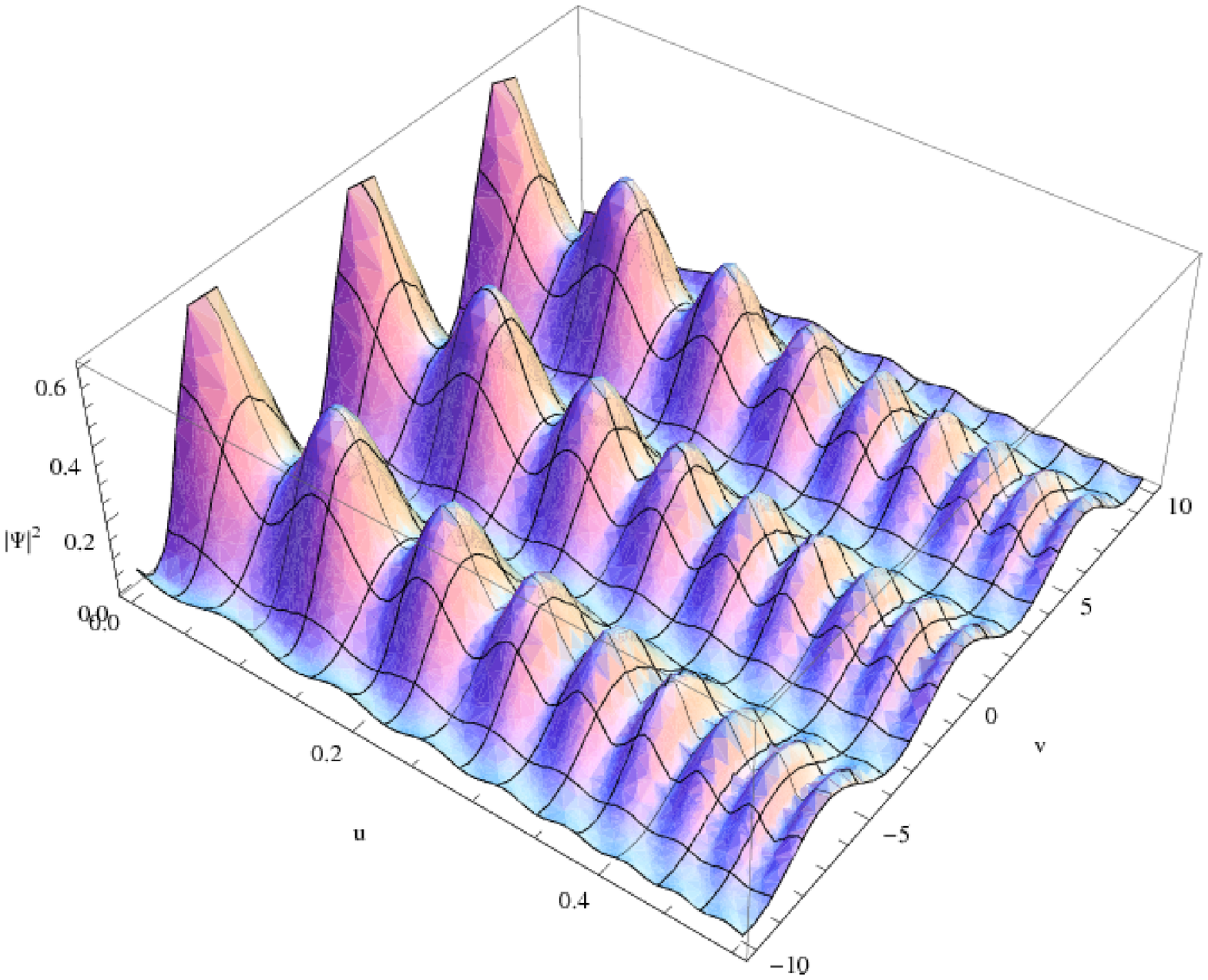,width=7cm}
\hspace{1cm} \epsfig{figure=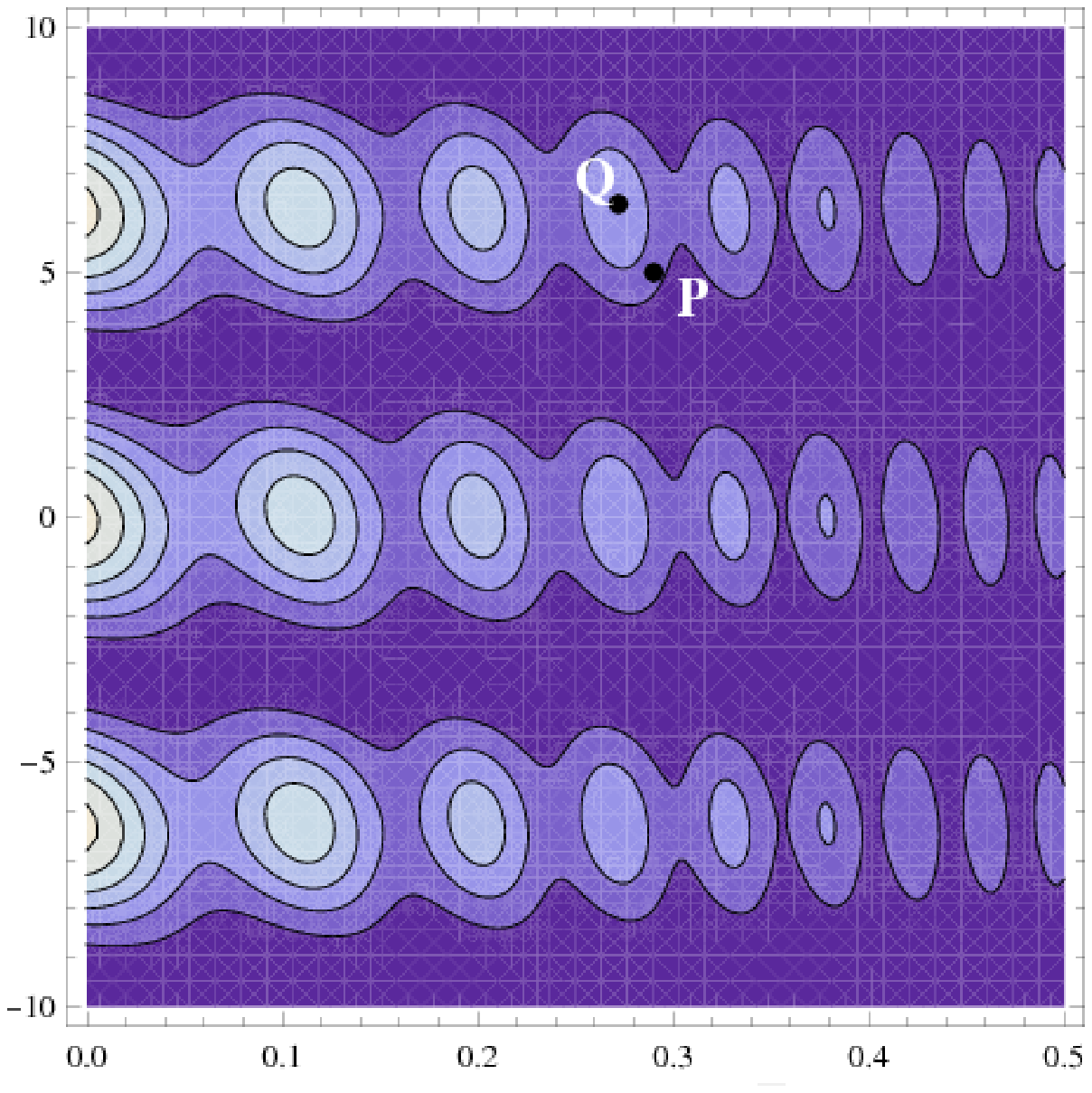,width=6cm}
\end{tabular}
\caption{\label{bianchi3}\footnotesize The right plot is shown
$|\Psi(u,v)|^2$ and the left is its corresponding contour plot. It
is for de Sitter case with $M_{\gamma}=20$. It is taken only two
first terms of relation (\ref{Bi14}) with a constant wight
function.}
\end{figure}

\begin{figure}
\begin{tabular}{ccc} \epsfig{figure=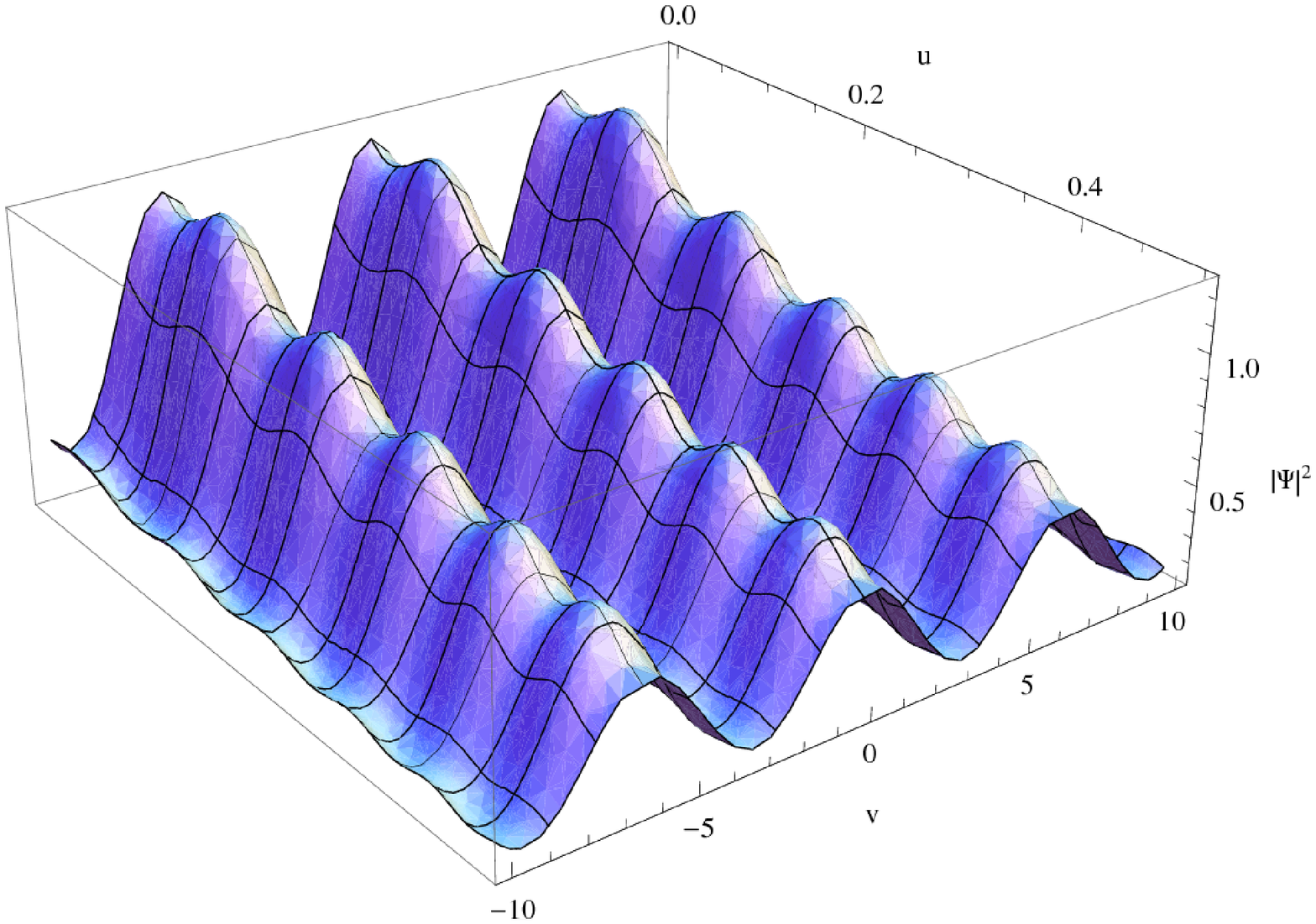,width=7cm}
\hspace{1cm} \epsfig{figure=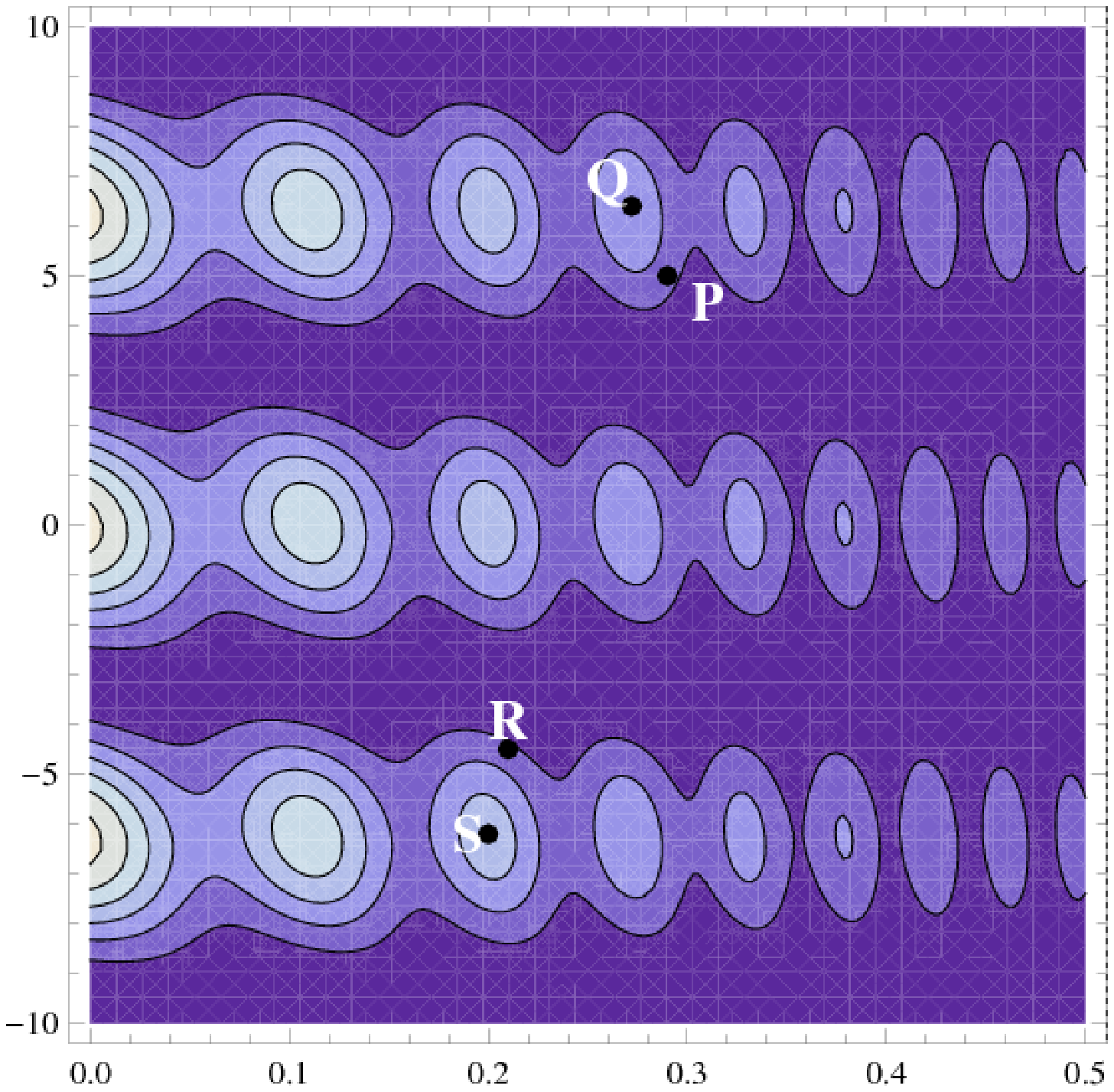,width=6cm}
\end{tabular}
\caption{\label{bianchi4}\footnotesize The right plot is shown
$|\Psi(u,v)|^2$ and the left is its corresponding contour plot. It
is for dust with $M_{\gamma}=20$. It is taken only two first terms
of relation (\ref{Bi14}) with a constant wight function.}
\end{figure}

The WD equation for our model
reads
\begin{equation}\label{Bi11}
{\cal H}\Psi(u,v)=0\Rightarrow \left[\frac{\partial^2}{\partial
u^2}-\frac{\partial^2}{\partial v^2}+24
M_{\gamma}e^{3(1-\gamma)u}\right]\Psi(u,v)=0.
\end{equation}
The solutions to the above equation are
\begin{equation}\label{Bi12}
\Psi_{\nu}(u,v)=e^{\pm i \nu v}J_{\pm i
\nu/3}(4\sqrt{\frac{M_{\gamma}}{6}}e^{3u}),\end{equation}for
$\gamma=-1$ and
\begin{equation}\label{Bi13}
\Psi_{\nu}(u,v)=e^{\pm i \nu v}J_{\pm2 i \nu/3}(4
\sqrt{\frac{2M_{\gamma}}{3}}e^{3u/2}),\end{equation} for $\gamma=0$,
where $\nu$ is a separation constant.  We have chosen oscillatory
function $e^{\pm i \nu v}$ since the real exponents would lead to
exponential increasing wave function for $v\rightarrow \pm \infty$
that would not show physical behavior. Therefore, we may now write
the general solution of the WD equation as a superposition of the
above eigenfunctions
\begin{equation}\label{Bi14}
\Psi(u,v)=\int_{-\infty}^{+\infty}C(\nu)\Psi_{\nu}(u,v)d\nu,
\end{equation}
where $C(\nu)$ can be chosen as a shifted Gaussian weight function
$e^{-a(\nu-b)^2}$. The square of wave function for different cases
are shown in figures \ref{bianchi3} and \ref{bianchi4} for de Sitter
case and dust respectively.
In figure \ref{bianchi3} point $Q$ is located at a local maximum
probability so PEP predicts a transition from point $P$ to $Q$ i.e.
$P{\buildrel {PEP} \over \longrightarrow }Q$. Physically, this
transition imposes going $u$ down and going $v$ up and after
$P{\buildrel {PEP} \over \longrightarrow }Q$ since $Q$ is located at
a maximum then $Q{\buildrel {PEP} \over \longrightarrow }Q$ that
means the system be steady at $Q$. This behavior is exactly similar
to previous result which is represented in the left plot in figure
\ref{ch1}. The similar is true for the left plot in figure \ref{ch1}
and $P{\buildrel {PEP} \over \longrightarrow }Q$ transition in
figure \ref{bianchi4}. Since point $S$ is located at a local maximum
probability so PEP predicts $R{\buildrel {PEP} \over \longrightarrow
}S$ then $S{\buildrel {PEP} \over \longrightarrow }S$. It means
there is this possibility that both of $u$ and $v$ go down. This
later result is completely in agreement with the previous subsection
results i.e. the represented case in figure \ref{bianchi2}.

There is a notable difference between Bianchi model and previous
ones that is existence of more than one direction in configuration
space for Bianchi model. This makes some ambiguities in path of
transition to more probable state in PEP paradigm. But as mentioned
roughly in \cite{nimagrg} the path in these kinds of examples is on
the gradient of $|\Psi|^2$ (hyper-)surface which has the most slope.
This proposal results in definition of a unique transition path in
PEP for more complicated configuration spaces.

\section{Discussion}\label{4}
In this paper we have argued, using familiar examples, that one can
draw a completely equivalent physical interpretation for two
different approaches to quantization, namely the usual canonical
quantization method and the phase-space deformation method, at least
within the framework of the examples used in this work. A by-product
is that when one studies the deformation of phase-space, one should
realize that this is effectively a quantization procedure and should
refrain from using any other quantization method simultaneously.
However, since in the phase-space deformation method for
quantization the evolutionary parameter, $t$, does appear, in
contrast to the canonical method, a mechanism for retrieving the
dynamical information of the system, when canonically quantized, is
at hand and can be used. This had been done in a previous work
\cite{nimagrg,grgg}. The results show that if the equivalence of
different methods of quantization discussed above are assumed, then
the probabilistic evolutionary mechanism works consistently and
therefore can be used to address the problem of evolution in
diffeomorphism invariant theories.

However, such an equivalence cannot hold true in examples where the
deformation in phase-space is introduced in a Lorentz non-invariant
manner. This is not hard to predict since the WD equation is a
direct result of diffeomorphism invariance and so if a deformation
in phase-space breaks such an invariance then the results of
different quantization methods should be different. In this
connection we note that $\kappa$-Minkowski deformation and GUP, for
example, preserve Lorentz invariance \cite{17}.

Another feature which is of interest is that the WD equation is
usually a second order differential equation and has therefore two
independent solutions. However, in the deformed phase-space quantum
models, the equations of motion are first order, in contrast to the
former. The disparity lies in the fact that, as discussed above, the
parameter of deformation, $\ell$, is arbitrary, and hence can be
chosen to be either positive or negative. It has been shown that
different signs of $\ell$ relate to different solutions of the WD
equation.

It is worth mentioning that this paper can be read as a companion
for our previous one \cite{nimagrg}. In \cite{nimagrg} we have
discussed on the PEP paradigm massively but here to make the current
manuscript self-studied we highlight some crucial features of PEP
very rapidly. The PEP proposal has some roots in the second law of
thermodynamics that is a system in an initial state evolves
approaching to another state in which the final state has a more
entropy value or a same one i.e. $S_f-S_i=\bigtriangleup S\geq 0$. A
vital problem in PEP is the velocity of transition from a state to a
more probable state. Actually, this question has not any concrete
answer yet due to the first steps in understanding PEP. But in a
model discussed in \cite{nimagrg} there is a relation for the
velocity of transition what is achieved by comparison of PEP's
results with Causal Dynamical Triangulation method as a model of
quantum gravity. It means until now it is an open problem that how
the velocity of transition can be addressed by PEP itself.

Finally, it is significant to consider that if a proposal is correct
in few examples it does not mean its correctness universally and in
this paper it is not our claim in any senses. But we have shown that
there is a kind of correspondence between different methods of
quantizations. We suppose, at least, this correspondence should be
considered until a complete understanding of the
universe\footnote{Specially for our examples, a theory of quantum
gravity or quantum cosmology.}. Since in lack of a full theory, e.g.
quantum gravity, all the approaches are considerable and perhaps
each approaches sheds some lights on various parts of the universe.
This viewpoint results in a proposal that a combination of different
approaches is a part of full theory\footnote{For example it is
believed in some communities that a combination of string theory and
loop quantum gravity can be an alternative for full quantum
gravity.}. However, maybe, different approaches have an intersection
in their territory of validness that have to be characterized
exactly to do not result in double consideration, e.g. double
calculations. In this paper we have shown that two different methods
of quantizations can be same as each other at least in some specific
examples and it must be considered to prevent double quantization.
And even more this correspondence can shed lights by each methods on
the other methods for example it can be interpreted as an evidence
for correctness of PEP method as a developing in understanding
canonical quantization method.

\end{document}